\documentclass[reprint,twocolumn,superscriptaddress,showpacs,nofootinbib,notitlepage]{revtex4-1}

\usepackage{graphicx}
\usepackage{latexsym,amsmath,amssymb,lmodern,float,url}
\usepackage{natbib}
\usepackage{color}
\usepackage{microtype}
\usepackage{slashed}
\usepackage{multirow}
\usepackage{comment}

\usepackage[colorlinks=true,backref=false, linktocpage=true,
citecolor=blue,urlcolor=blue,linkcolor=blue,pdfpagemode=UseOutlines]{hyperref}

\hypersetup{%
  bookmarksnumbered=true,
  pdftitle = {},
  pdfsubject = {},
  pdfauthor = {},
  pdfkeywords = {}
}

\let\Re\undefined

\DeclareMathOperator{\Re}{Re}

\renewcommand{\d}{\mathrm d}

\begin{document}
\title{Parton Physics on a Quantum Computer}
\author{Henry Lamm}
\email{hlamm@umd.edu}
\affiliation{Department of Physics, University of Maryland, College Park, Maryland 20742, USA}
\author{Scott Lawrence}
\email{srl@umd.edu}
\affiliation{Department of Physics, University of Maryland, College Park, Maryland 20742, USA}
\author{Yukari Yamauchi}
\email{yyukari@umd.edu}
\affiliation{Department of Physics, University of Maryland, College Park, Maryland 20742, USA}
\date{\today}

\collaboration{NuQS Collaboration}
\begin{abstract}
Parton distribution functions and hadronic tensors may be computed on a universal quantum computer without many of the complexities that apply to Euclidean lattice calculations. We detail algorithms for computing predictions of parton distribution functions and the hadronic tensor in the Thirring model. Their generalization to QCD is discussed, with the conclusion that the parton distribution function is best obtained by fitting the hadronic tensor, rather than direct calculation.  As a side effect of this method, we find that lepton-hadron cross sections may be computed relatively cheaply. Finally, we estimate the computational cost of performing such a calculation on a digital quantum computer, including the cost of state preparation, for physically relevant parameters.
\end{abstract}

\maketitle 
\section{Introduction}

Parton distribution functions (PDFs) and hadronic tensors have been widely studied as they compactly parameterize the hadronic structure as seen by high-energy probes. They provide the nonperturbative input to deep inelastic scattering cross sections, as well as the initial conditions for heavy ion experiments. As a result of this fundamental importance, significant effort has been expended on both the experimental and theoretical side~\cite{PhysRevD.98.030001}.

As the PDF is a non-perturbative object, existing first-principles calculations proceed from the Euclidean lattice \cite{Liu:1993cv,Aglietti:1998mz,Liu:1999ak,Ji:2013dva,Chambers:2017dov,Bali:2018spj,Radyushkin:2017cyf,Orginos:2017kos,Karpie:2018zaz,Karpie:2019eiq,Cichy:2019ebf,Alexandrou:2019lfo,Sufian:2019bol,Ma:2014jla,Ma:2014jga,Ma:2017pxb,Liang:2019frk}. This presents a fundamental difficulty: the PDF is defined as an integral of real-time separated correlators, and direct computation on the lattice is prevented by a sign problem. Several approaches have been pursued to circumvent this difficulty, including analytic continuation~\cite{Aglietti:1998mz,Liu:1999ak}, computing the hadronic tensor~\cite{Ma:2014jla,Ma:2014jga,Ma:2017pxb,Liang:2019frk}, Compton amplitudes~\cite{Chambers:2017dov}, and quasi-PDFs~\cite{Ji:2013dva} or pseudo-PDFs~\cite{Radyushkin:2017cyf}, each of which carries unique technical complications.

The near-advent of small-scale universal quantum computers has stimulated research into their possible applications. Of particular interest here are the algorithms for achieving real-time evolution of a field theory, particularly gauge theories~\cite{Banerjee:2012pg,Hauke:2013jga,Kuno:2014npa,Kuno:2016ipi,Martinez:2016yna,Marcos:2014lda,Klco:2018kyo,Zache:2018jbt,Muschik:2016tws,Bazavov:2015kka,Zhang:2018ufj,Unmuth-Yockey:2018xak,Unmuth-Yockey:2018ugm,Gustafson:2019mpk,Zohar:2012ay,Byrnes:2005qx,Banerjee:2012xg,Wiese:2013uua,Zohar:2012xf,Tagliacozzo:2012df,Zohar:2016iic,Bender:2018rdp,Lamm:2019bik,Alexandru:2019nsa,Klco:2019evd,Davoudi:2019bhy}. Classically hard, real-time evolution is among the most natural operations available on a quantum computer. However, it is unclear exactly what quantities of physical interest may be obtained more easily by this method. Inclusive cross sections, for instance, are a natural target~\cite{Jordan:2011ci,Jordan:2014tma,Jordan:2017lea}, but are already in practice obtainable classically~\cite{Luscher:1990ux}. In this paper we propose that PDFs and the hadronic tensor may be a \emph{physically relevant} target on which ``quantum supremacy'' --- that is, the ability of quantum computers to calculate quantities unobtainable classically --- may be demonstrated. Additionally, this provides a cheaper way to compute lepton-hadron cross sections than that discussed in~\cite{Jordan:2011ci,Jordan:2014tma,Jordan:2017lea}. Finally, these methods could complement Euclidean lattice methods to obtain theoretical predictions in partonic physics with different systematics.

The largest quantum computers currently available are limited to tens of qubits, and in the near future, computers with more than a few hundred, moderately noisy qubits cannot be reasonably expected. This situation has been termed the NISQ (Noisy, Intermediate-Scale Quantum) era.  With an eye toward calculations achievable in the NISQ era, \cite{Mueller:2019qqj} proposed a method for computing the hadronic tensor, in the Regge limit, in the framework of color glass condensate EFT. In this work we discuss a more expensive prospect: first-principles calculations in the framework of Hamiltonian lattice field theory.

The method discussed here, as applied to $3+1$ dimensional QCD, requires computers several orders of magnitude larger than anything expected in the NISQ era.  In contrast, the $1+1$ dimensional Thirring model we discuss requires substantially fewer resources and could plausibly be studied in the NISQ era as a toy model. Our purpose is not to present new calculations, but to describe how these observables can be computed and what sort of resources are required to obtain, on a quantum computer, these particular physical observables from first principles. To that end, we analyze the cost of adiabatic state preparation of the proton, as well as the practical cost of time-evolution on a sufficiently sized lattice.

For illustration purposes we work with the staggered Thirring model in one spatial dimension. The Hamiltonian of this model is
\begin{align}\label{eq:hamiltonian}
H = \sum_x
&\;\frac 1 2 \left(-1\right)^x
\left(\chi^\dagger(x) \chi(x+1) + \chi^\dagger(x+1) \chi(x) \right)
\nonumber\\
&+ m \left(-1\right)^x \chi^\dagger(x) \chi(x)\nonumber\\
&-g^2 \chi^\dagger(x) \chi(x) \chi(x+1) \chi^\dagger(x+1)
\end{align}
where $\chi$ denotes a single-component fermion field, which becomes a two-component field in the continuum limit. The Hilbert space of this model is small enough to allow us to simulate the evolution of the system with classical resources on small lattices. The Thirring models differs importantly from QCD in not being a gauge theory; we will see that for the evaluation of the hadronic tensor this makes no significant difference to the method, but direct computations of PDFs are dramatically affected.

The rest of this paper is structured as follows. First we describe the direct computation of the PDF in Sec.~\ref{sec:pdf}. As we shall see, this is entirely impractical in a gauge theory, and so in Sec.~\ref{sec:ht} we show how the hadronic tensor is obtained instead. A theoretical prediction for the PDF may be extracted from the hadronic tensor via the same procedure used by experiments. In Sec.~\ref{sec:state} we discuss the preparation of the ground state of a proton --- a prerequisite for any of these procedures. We conclude in Sec.~\ref{sec:discussion} with a discussion of the future prospects for this and similar methods.

\section{Quark Distributions}\label{sec:pdf}

Central to hadronic physics are the quark and gluon distribution functions~\cite{Collins:1981uw}. The distribution functions $f(x)$ may be interpreted as giving the probability for a high-energy probe to see a parton with a given momentum $xP$ within a hadron with momentum $P$. It is most natural to consider $f(x)$ on the light cone; however, in this paper we will view it in equal-time quantization, as that is the framework in which we would simulated a field theory on a quantum computer.

We use the Thirring model for illustration, and by analogy with QCD we refer to the fundamental fermion as a `quark' and the bosonic bound state as a `meson'. Note that, because the Thirring model is not confining, the dressed fermion is itself an asymptotic state, of which a PDF may be computed. The quark distribution function in the Thirring model is simpler than that of QCD because no Wilson line is needed. It is given by
\begin{equation}
f(x) = \int \d y \;e^{i x P^+ y}
\left<
P
\right|
\bar\psi(y) \gamma^+ \psi(0)
\left|
P
\right>
\text,
\end{equation}
where $\gamma^+ \equiv \frac 1 {\sqrt 2} (\gamma^0 + \gamma^1)$, and $P^+ \equiv \frac 1 {\sqrt 2} (P^0 + P^1)$ is the lightcone momentum of the incoming hadron. The expectation value is taken in the ground state of a hadron with momentum $P$.
Here we see that the PDF is the Fourier transform of a time-separated correlator.

Our strategy will be to calculate the integrand on a quantum computer for many values of $y$, and then approximate the Fourier transform classically. In a form more readily obtained on a quantum computer, the quark correlator is
\begin{equation}
\phi(y) = \left<P\right|e^{i H y^0} \bar\psi(\vec y) e^{-i H y^0} \gamma^+ \psi(0) \left|P\right>
\text.
\end{equation}
The preparation of the hadron state $\left|P\right>$ is involved, and we postpone its discussion to Sec.~\ref{sec:state}. Assuming that $\left|P\right>$ is readily prepared, we must now translate this expression to the lattice, where $\psi$ is staggered and $y$ is discrete. It may be seen by taking the Fourier transforms of the fields $\psi$ and $\chi$ that an appropriate ``staggered PDF'' is
\begin{align}
f_{\mathrm{stag}}(x) = 
\left<P\right|
\sum_{y,z} &e^{ixP(y-z)}
\left[
\delta^{|y|}_{|z|}
+
i (-1)^z \delta^{|y|}_{|z+1|}
\right]
\nonumber\\\times &
e^{i H (y-z)} \chi^\dagger(y) e^{-i H (y-z)} \chi(z)
\left|P\right>\label{fstag}
\text.
\end{align}
For brevity, we have used the notation $|y| = 0$ when $y$ is even and $1$ when $y$ is odd.

We must translate the operators $\chi^\dagger$, $\chi$, and $e^{-i H t}$ into bosonic qubit operators. The operators $\chi(y)$ and $\chi^\dagger(y)$ are anticommuting, and are constructed from bosonic qubit operators $\sigma$ via the Jordan-Wigner transformation~\cite{Jordan:1928wi}. The construction of $e^{-i H t}$ follows the standard techniques of Trotterization. The Hamiltonian is split up into several mutually non-commuting terms $H_x,H_y,$ and $H_z$, each of which is easily diagonalized in isolation. A single time step under $H$ is then approximated by a brief period of time-evolution under each of the terms alternately: $e^{-i (H_x+H_y+H_z) \delta} \approx e^{-i H_x \delta} e^{-i H_y \delta}e^{-i H_z \delta}$.  After the Jordan-Wigner transformation, the three terms of the Hamiltonian are
\begin{eqnarray}
H_x &=& \sum_{n=1}^{N} a_{xx}(n)\sigma_x(n)\sigma_x(n+1)\nonumber\\
H_y &=& \sum_{n=1}^N a_{yy}(n)\sigma_y(n)\sigma_y(n+1)\text{, and}\nonumber\\
H_z &=& \sum_{n=1}^{N} a_z(n)\sigma_z(n) + a_{zz}(n)\sigma_z(n)\sigma_z(n+1)\text.
\end{eqnarray}
where $a_z(n)=m(-1)^{n}/2$, $a_{xx}(n)=a_{yy}(n) = (-1)^{n+1}/4$, $a_{zz}(n)=g^2/4$.

The operator that forms the matrix element in Eq.~(\ref{fstag}) is not Hermitian.  In order to evaluate this non-Hermitian operator on a quantum computer, we decompose into a sum of unitary operators, each of which may be evaluated \`a la \cite{Ortiz:2000gc} with the help of an ancillary qubit. The decomposition of the operator in Eq.~(\ref{fstag}) is:
\begin{eqnarray}
 &&e^{i H (y-z)}\chi(y)e^{-i H (y-z)}\chi^{\dag}(z)= \sum_{i,j=x,y} C_{ij}  U_{ij}  \label{unitary}\\
 && U_{ij} = e^{i H (y-z)}\chi_i(y)e^{-i H (y-z)}\chi_j(z)
\end{eqnarray}
where $\chi_x = \chi + \chi^\dagger$ and $\chi_y = i(\chi - \chi^\dagger)$.
The coefficients $C_{ij}$ are determined from the Jordan-Wigner transformation: $C_{xx}=1/4, C_{xy}=-i/4,C_{yx}=i/4,C_{yy}=1/4$. Following \cite{Ortiz:2000gc}, each term in Eq.~(\ref{unitary}) is measured on a quantum computer by first preparing the state
\begin{equation}
|P'\rangle = \frac{1}{\sqrt{2}}\left(  |0\rangle_a |P\rangle + |1\rangle_a |P\rangle \right),
\end{equation}
where $a$ denotes the ancillary qubit, and then applying $U_{ij}$ controlled on the ancillary to $|P'\rangle$. Measurements of $\sigma_x$ and $\sigma_y$ of the ancillary qubit on the resulting state give us the real and imaginary parts, respectively, of the terms in Eq.~(\ref{unitary}), $\langle P | U_{ij} | P \rangle$.

Two technical complications remain. First, because the correlators are evaluated along the light-cone the speed of light must be known precisely. Without the hypercubic symmetry of the Euclidean lattice preventing renormalization, the speed of light must be computed nonperturbatively. In principle, we could measure the speed of light on the quantum computer. This is a formidable task, entailing careful measurements of the dispersion relation near the continuum limit. But this is in fact unnecessary. The dispersion relation is reflected in the low energy portion of the spectrum of the Hamiltonian, which can be readily determined on an anisotropic Euclidean lattice~\cite{Engels:1981qx,Burgers:1987mb}. Thus the speed of light on a quantum computer may be determined without any calculations being performed on a quantum computer, as long as the Hamiltonian limit is taken on both the classical and quantum machines.  In the specific case at hand of the $1+1$ dimensional Thirring model, the situation is even simpler: numerical experiments reveal that the speed of light in the continuum limit is $1$ in lattice units.

The speed of light is only defined in the continuum limit. On the lattice, no exact light-cone exists, and `space-like' separated fermionic operators need not exactly anticommute. As a result, the lattice PDF will not have the desired symmetry properties until the continuum limit is taken. Additionally, if periodic boundary conditions are used, care must be taken not to evaluate the quark correlator at separations larger than the spatial size of the lattice.

Finally we must take the Fourier transform. Only a finite number of values of the quark correlator may be computed, and naively taking the Fourier transform will show highly oscillatory artifacts (as in Euclidean lattice calculations~\cite{Karpie:2019eiq}). In order to take the Fourier transform in a stable way, without these artifacts, we impose a Gaussian window, defining
\begin{equation}\label{eq:ft-reg}
f(x) = \int_{-L}^L \d x \;e^{i x P^+ y - x^2 / \epsilon}
\phi(y)
\text,
\end{equation}
and first taking the limit $L \rightarrow\infty$, and only then allowing $\epsilon\rightarrow 0$. These limits may be taken numerically.

\begin{figure}
\centering
\includegraphics[width=\linewidth]{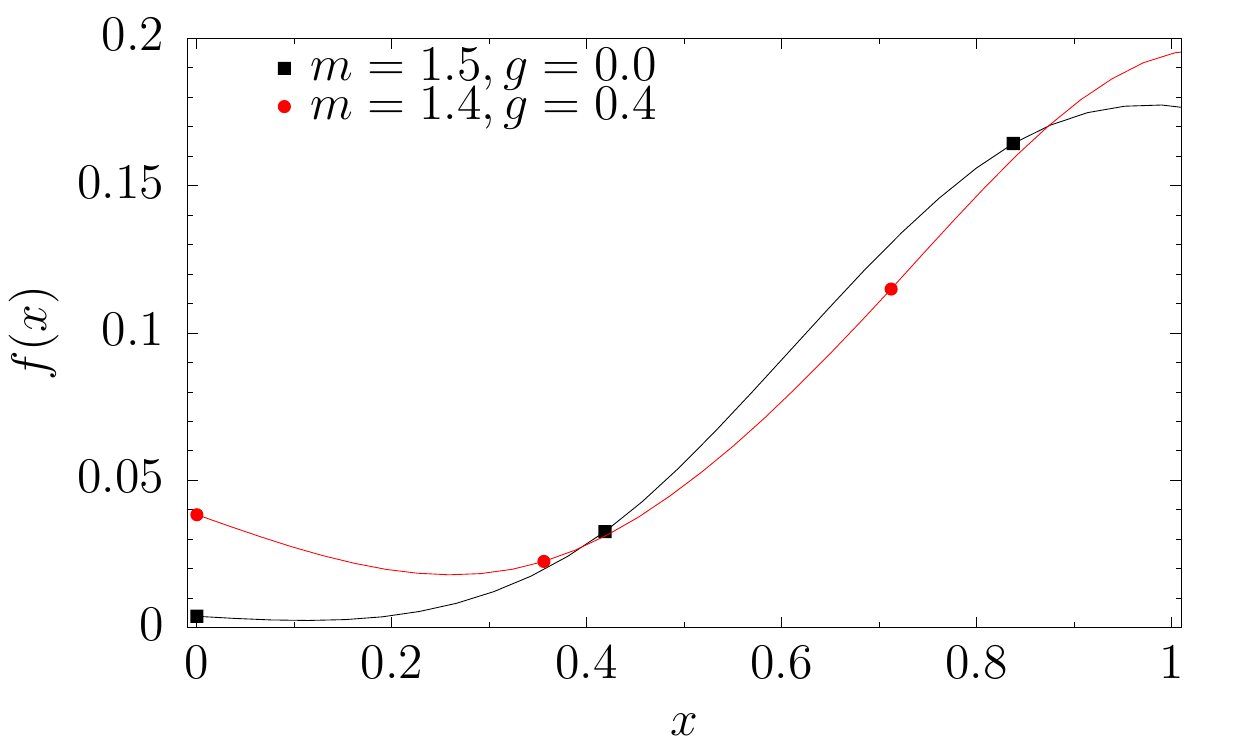}
\caption{The quark distribution function of the lowest-lying fermion in the Thirring model, computed on a $10$-site lattice. The Fourier transform of Eq.~(\ref{eq:ft-reg}) is taken with $\epsilon=3$.\label{fig:pdf}}
\end{figure}

This completes the description of how to obtain the PDF of the Thirring model on a quantum computer, given an already-prepared hadronic state. In Fig.~\ref{fig:pdf} is shown a calculation, by exact diagonalization of the Hamiltonian, of the fermion distribution function of the lowest-lying fermion state at vanishing and weak couplings.

The PDFs of asymptotic states in the Thirring model may be interesting in the context of quantum supremacy: the qubit cost of this calculation makes it potentially accessibly in the NISQ era, while classical algorithms struggle to obtain this observable. Ultimately, however, we would like to calculate the PDFs of mesons and baryons in QCD. Quark distribution functions in QCD are given by
\begin{equation}\label{eq:qcd-pdf}
f_i(x) = 
 \int \d x \;e^{i x P^+ y}
\left<
P
\right|
\bar\psi_i(y) \gamma^+ W(y;0) \psi_i(0)
\left|
P
\right>
\end{equation}
where $W(y;0)$ is a light-like Wilson line connect $y$ to the origin required to ensure gauge-invariance~\cite{Collins:1981uw}, and $i$ enumerates the quark flavors.

Several proposals have been advanced for how gauge field theories may be simulated on a quantum computer. Let us consider the scheme laid out in \cite{Lamm:2019bik}, which provides a procedure for computing Wilson loops. In this scheme, time-evolution is implicitly performed in $A_0 = 0$ gauge, so that a $W(y;0)$ is approximated by a sequence of spatial link operators applied at different points in time. (The time-like links are fixed to be the identity in this gauge.)
\begin{equation}
W(y;0) \approx e^{i H y} W(y;y-a) e^{-i H a} \cdots e^{-i H a} W(a;0)
\end{equation}
In \cite{Lamm:2019bik} it was shown that obtaining a time-separated correlator of two gauge links (i.e. a temporal Wilson loop) requires a second-order derivative to be taken numerically. Here, perturbations to the Hamiltonian occur at every time slice between the two operators (necessarily, so that the Wilson line is approximately light-like). The order of the finite differencing needed is equal to the number of time slices affected. This high-order finite differencing is not practical even in the absence of quantum noise. Nevertheless, this is the only candidate we are aware of for directly computing correlators of the form Eq.~(\ref{eq:qcd-pdf}) on a quantum computer.

Fortunately an alternative procedure can be constructed: one may compute an easier observable --- the hadronic tensor --- and extract the PDFs after the fact.

\section{Hadronic Tensor}\label{sec:ht}
Unlike the PDF, the hadronic tensor is constructed of currents, $J^\mu$, which are each gauge-invariant, unlike the fermionic operators.  Thus, the hadronic tensor does not require a Wilson line and the issue of high-order finite differencing is avoided.
The hadronic tensor of a $d$-dimensional theory is given explicitly by
\begin{equation}\label{eq:ht}
W^{\mu\nu}(q) = \Re \int \d^d x\;e^{i q x} \left<P\right|T \left\{J^\mu(x) J^\nu(0)\right\}\left|P\right>
\end{equation}
for a given current $J^\mu$, where $\left|P\right>$ denotes a proton in the zero momentum frame.  Here we will assume $J^\mu = \bar\psi \gamma^\mu \psi$, corresponding to the current coupling to the photon. In combination with the leptonic tensor $L_{\mu\nu}$,
\begin{equation}
L_{\mu\nu} = 2\left( k_{\mu}k'_{\nu} + k_{\nu}k'_{\mu} -g_{\mu\nu}k\cdot k' \right)\text,
\end{equation}
$W^{\mu\nu}(q)$ may be directly related at leading order (in the QED coupling $\alpha$) to the cross section of lepton-proton scattering via the exchange of a photon with momentum $q$ by~\cite{PhysRevD.98.030001}
\begin{equation}
\frac{d^2\sigma}{dxdy} = \frac{\alpha^2 y}{Q^4}  L_{\mu\nu} W^{\mu\nu}
\text.
\end{equation}
In these equations, $Q^2 = -q^2$, $x = Q^2/2P\cdot q$, $y = P\cdot q/P\cdot k$, and $k' = k-q$. 

With an eye toward implementation on a quantum computer, Eq.~(\ref{eq:ht}) has two critical features. First, it is gauge-invariant without the need for a Wilson line. In fact, because it involves only gauge-invariant operators, it can be defined and measured without reference to unphysical gauge-variant (Gauss-law-violating) states. Second, each operator in the correlator is individually Hermitian, so that the decomposition procedure of the previous section is not required.

Several options exist for measuring the integrand of the hadronic tensor. One may follow the procedure of the previous section closely, decomposing $J^\mu(x)$ as a sum of $N$ unitary operators. After this, the operator in the integrand becomes a sum of $N^2$ unitary operators, each of whose expectation values may be directly evaluated \` a la \cite{Ortiz:2000gc}. However, as mentioned above, this is needlessly complicated for the hadronic tensor.

In this context, the procedure of \cite{PhysRevLett.113.020505} for studying linear response is more straightforward. Consider a unitary evolution operator
\begin{equation}
U(\epsilon_x,\epsilon_0)
=
e^{i H t} e^{i J^{\mu}(\vec x) \epsilon_x}e^{-i H t} e^{-i J^{\nu}(\vec 0) \epsilon_0}
\text.
\end{equation}
The first derivative of the expectation value of this operator with respect to either $\epsilon$ vanishes. The second derivative gives the desired correlator.
\begin{equation}\label{eq:integrand}
\frac{\d}{\d\epsilon_x}
\frac{\d}{\d\epsilon_0}
\left<P\right|
U(\epsilon_x, \epsilon_0)
\left|P\right>
= \left<P\right|J^\mu(x)J^\nu(0)\left|P\right>
\end{equation}

Finally, a more sophisticated procedure for the calculation of linear response is given in \cite{Roggero:2018hrn}.

After measuring Eq.~(\ref{eq:integrand}) at many values of $x$, the Fourier transform may be taken classically, via Eq.~(\ref{eq:ft-reg}). Alternatively, the (regulated) Fourier transform may be subsumed into the expectation value, yielding
\begin{equation}\label{eq:ht-phys}
W^{\mu\nu}(q) = \left<P\right|T \left\{J^\mu(0) \int \d^d x \; e^{i q x - x^2/\epsilon} J^\nu(x)\right\}\left|P\right>
\text.
\end{equation}
This expression requires only one quantum circuit for a desired value of $q$; however, when many values of $q$ are to be obtained, it is no more efficient. 

There is an important way in which these procedures are not analogous to those performed in the laboratory: on a quantum computer, one may introduce a current coupling only to particular flavors of fermions. This allows one to isolate a single-flavor distribution function or hadronic tensor in a straightforward way, without any fitting.

To apply this method to the Thirring model, one needs the staggered form of $J^\mu$:
\begin{eqnarray}
J^0(x) &=& \chi^\dagger(x) \chi(x)\\
J^1(x) &=& \frac i 4 (-1)^x \big[\chi^\dagger(x) \left(\chi(x+1) + \chi(x-1)\right) \nonumber\\ & & -\left(\chi^\dagger(x+1) + \chi^\dagger(x-1)\right) \chi(x)\big]
\text.
\end{eqnarray}

As mentioned, the leading order cross section for lepton-hadron scattering may be computed once the hadronic tensor is in hand. This is not the first proposal for computing a scattering cross-section on a quantum computer; in \cite{Jordan:2011ci,Jordan:2014tma,Jordan:2017lea} is detailed a procedure in which two asymptotic states are prepared adiabatically on a large lattice, and then allowed to propagate towards each other. When obtaining a cross section via the hadronic tensor, the need to prepare two asymptotic states is removed-- reducing substantially the cost of state preparation. Instead, we prepare only a single asymptotic (zero-momentum, in fact) state, and probe it with arbitrary momentum.  Additionally, this avoids the long-range nature of the QED interaction that complicates lattice calculations. This procedure is substantially simpler, but the tradeoff comes in that while~\cite{Jordan:2011ci,Jordan:2014tma,Jordan:2017lea} computes the full cross section, our procedure is perturbative: to obtain higher-order contributions in $\alpha$, one must calculate multiple matrix elements defined by additional current insertions.

If one ultimately wants the PDF, one can extract it from $W^{\mu\nu}(q)$ via a procedure analogous to how the experimental determinations are done.  To first review, there are a number of processes where \emph{collinear factorization} can be proven (e.g. deep inelastic scattering, Drell-Yan, weak boson production, and inclusive jet production).  Here we consider deep inelastic scattering, but similar expressions are derivable for the other processes.  The cross section $\sigma_{eP\rightarrow eX}$ can be schematically decomposed into 
\begin{equation}
\label{eq:dis}
 \sigma_{eP\rightarrow eX}=\sum_{i,j}f_i\otimes P_{i\rightarrow j}\otimes \sigma_{ej\rightarrow ej}
\end{equation}
where $i$ and $j$ run over all species of parton, $f_i$ are parton distributions, $P_{i\rightarrow j}$ is the splitting function required to match all experimental data at a single scale and can be computed perturbatively, and $\sigma_{ej\rightarrow ej}$ is the hard partonic cross section.  Theoretical expressions like Eq.~(\ref{eq:dis}) are used to numerically fit parameterized PDFs to the experimental data over large ranges of kinematics~\cite{PhysRevD.98.030001}.  The complicated nature of the perturbative splitting function and hard cross section in addition to the need to perform two convolutions prove to make this process highly nontrivial.

In the same spirit, the hadronic tensor that would be obtained by a quantum computer can be defined in terms of the PDFs as
\begin{equation}
\label{eq:hdpdf}
 W^{\mu\nu}=\sum_{i,j}f_i\otimes P_{i\rightarrow j}\otimes\hat{W}^{\mu\nu}
\end{equation}
where $\hat{W}^{\mu\nu}$ are partonic tensors that couple to external currents.  Thus, if one desires the PDFs, they can be extracted by numerical fits to parameterized PDFs when the hadronic tensor is computed in the kinematic regime of collinear factorization's validity.  It is important to note that our procedure can, by allowing different four-momentum for the hadronic states, be trivially generalized to computing Generalized Parton Distributions~\cite{Belitsky:2012ch} and similar small changes to obtain other distributions--something that is not trivially possible for Euclidean field theory.

\section{State Preparation}\label{sec:state}
Thus far we have neglected to discuss the preparation of the state $\left|P\right>$. This is not a trivial matter, and in this section we rectify the situation. For concreteness, let us assume that we are preparing a zero-momentum proton on a three-dimensional QCD lattice.

A great deal of literature discusses the problem of preparing a ground state of a given Hamiltonian~\cite{Kaplan:2017ccd,Lamm:2018siq,Jordan:2011ci,Jordan:2014tma,Jordan:2017lea}. Typically, formal analysis of the efficiency of ground state preparation methods is not available, and the in-practice performance cannot yet be measured. Here we consider adiabatic state preparation \cite{Jordan:2011ci,Jordan:2014tma,Jordan:2017lea}, where we will be able to make some crude estimates of the cost of preparing a proton. Note that this does not imply adiabatic state preparation is the most efficient method in practice, simply that it is the easiest to analyze without the need to test on a full-scale quantum computer. Other possibilities for ground state preparation, not to be discussed here, are the spectral comb~\cite{Kaplan:2017ccd} and hybrid methods~\cite{Mueller:2019qqj,Lamm:2018siq}.

When a method is cast as preparation of a ground state, it is typically still applicable to the preparation of a state like $\left| P\right>$: the preparation of the proton may be translated to ground state preparation. To do this, consider only the sector of Hilbert space which is translation-invariant (thus zero-momentum) and has quantum numbers of the proton. As long as time evolution $e^{-i H t}$ maps this sector to itself (as standard Trotterized time evolution does), we may safely use the algorithm to prepare the ``ground state'' of this sector specifically, which is of course the zero-momentum proton. Note that this ``trick'' is exactly the same as what is needed to prepare a gauge-invariant ground state, where we must restrict ourselves to the physical subspace of the Hamiltonian.

We now estimate the costs of adiabatic state preparation in the context of QCD. We make several assumptions about the spectrum of the lattice model. In the restricted Hilbert space with baryon number $1$, the lowest-lying state should be the zero-momentum proton. If simulating pure QCD (that is, in the absence of weak interactions), we may further restrict to the states with isospin of the proton, thus avoiding a small gap between the proton and the neutron. Finally, on a finite volume the gap between a particle and a slowly moving particle is $\mathcal O(\frac 1 L)$. If we restrict to the zero-momentum subspace, then the gap between the proton and the nearest-energy state is the pion mass, $m_\pi \approx 135\;\mathrm{MeV}$.

Adiabatic state preparation begins by preparing the ground state for some modified Hamiltonian, for which the ground state is known with great precision. In this case we chose the free theory. The ground state of the baryon-number-$1$ sector is three zero-momentum fermions in a box. As the gauge coupling is $0$, the configuration of the gauge fields in the ground state is also Gaussian, and may be prepared efficiently.

Adiabatic state preparation proceeds by slowly deforming the Hamiltonian, over a time $T$, from the initial (in this case, free) Hamiltonian $H_0$ to the desired Hamiltonian $H_T$. The simplest trajectory we can pick increases the coupling from $0$ to its desired physical value at a constant rate. The deformation must be done slowly, and more slowly when the gap in $H_t$ is small. The adiabatic theorem guarantees we will remain in the ground state (with high probability) as long as $\dot H / \Delta^2 \ll 1$, where $\Delta$ is the gap and $\dot H$ is the rate of change of the Hamiltonian~\cite{messiah1962quantum}. Thus to estimate the performance of the adiabatic procedure we must estimate the gap along the trajectory. At the end of the trajectory, as mentioned, the gap is large: about $1/7$ the mass of the proton. This part of the evolution can be done quickly. At vanishing coupling, the outlook is less rosy. The `proton' fills the lattice, and excited states are simply back-to-back low-momentum states of two fermions. (The massless glue excitations can be removed with appropriate boundary conditions, or by using the $1080$-element approximation to $SU(3)$~\cite{Alexandru:2019nsa,Lamm:2019bik}.) The gap, therefore, is $\mathcal O(1/L)$, and we see that the adiabatic algorithm will require $\mathcal O(L^2)$ time-evolution steps to keep the ground state.

Not all ground-state preparation methods may have this scaling. As an example, spectral combing \cite{Kaplan:2017ccd} does not require a trajectory starting from weak coupling (thus avoiding the $1/L$ gap), and appears in numerical studies to scale like $\Delta^{-1}$. However, in the absence of large-scale tests, it is unclear how faithfully the method actually prepares a ground state.

\section{Discussion}\label{sec:discussion}

With the procedure completely described, we may estimate what manner of quantum resources are required for a practical calculation of the hadronic tensor of the proton. For the sake of specificity we will assume that the proposal of \cite{Alexandru:2019nsa,Lamm:2019bik} is used; namely, that $SU(3)$ gauge theory is approximated by the $1080$-element discrete subgroup, and the calculation is done in $A_0 = 0$ gauge with no further gauge fixing. The qubit cost of other methods is expected to be similar. The applicability of the $S(1080)$ approximation is limited by the lattice spacing at which the lattice theory undergoes a phase transition. It was shown in \cite{Alexandru:2019nsa} that it remains a good approximation down to a lattice spacing of $a=0.08\;\mathrm{fm}$, suggesting that it is sufficient for low- and intermediate-momentum probes of hadronic physics~\cite{Alexandrou:2019lfo}.

The qubit costs are easiest to count: representing a single element of the group $S(1080)$ on a link must cost at least $11$ qubits, and therefore an $L^3$ lattice requires, taking into account a $12$-component ($4$ spinor by $3$ color) Wilson fermion at each site, $\sim 50 L^3$ qubits to store. Time-evolution brings in the need for some number of ancillary qubits, but this need not scale with the volume of the system, and must therefore be negligible. Assuming a lattice spacing $a = 0.1\;\mathrm{fm}$ and a $20^3$ lattice --- chosen to be well within the range of applicability of $S(1080)$ and large enough to fit a proton with moderate finite-volume effects, $\sim 4 \times 10^5$ qubits are required to perform the calculation.

Estimating the gate cost associated with the calculation is difficult, not least because circuits have not yet been produced for any concrete proposal for simulating $SU(3)$ gauge theory. Nevertheless, we can give the scaling of the algorithm with volume. The procedure is dominated by state preparation, which requires time evolution for $\mathcal O(L^2)$ steps. In computing the hadronic tensor we must evaluate $L^3 \times T$ matrix elements, where $T$ gives the length of time evolution needed in approximating the Fourier transform. As the $J_\mu(x)$ at a single time are mutually commuting, $L^3$ measurements can be performed simultaneously, without needing to re-evolve the system. Assuming the evolution time $T$ to be proportional to $L$, we find that $\mathcal O(L^3)$ time evolution steps will be required. Each time evolution step, of course, scales with the volume of the lattice, so the total scaling of the procedure should be $\mathcal O(V^2)$. This is comparable to the scaling of calculations performed on the Euclidean lattice.

In this paper we have detailed a possible application for large-scale quantum computers beyond the NISQ era: first-principles calculations of parton physics in field theories, particularly QCD. As a side effect, this would allow the calculation of hadron-lepton cross sections in the standard model, more cheaply than existing proposals. Although not explored in this paper, a nearer-term target of this method is the physics of bound states in fewer dimensions.


\begin{acknowledgments}
Thanks are due to Evan Berkowitz, Tom Cohen, Aram Harrow, and Yao Ji for many helpful comments. H.L., S.L., and Y.Y. are supported by the U.S. Department of Energy under Contract No.~DE-FG02-93ER-40762.
\end{acknowledgments}
\bibliographystyle{apsrev4-1}
\bibliography{pdf}

\begin{thebibliography}{60}%
\makeatletter
\providecommand \@ifxundefined [1]{%
 \@ifx{#1\undefined}
}%
\providecommand \@ifnum [1]{%
 \ifnum #1\expandafter \@firstoftwo
 \else \expandafter \@secondoftwo
 \fi
}%
\providecommand \@ifx [1]{%
 \ifx #1\expandafter \@firstoftwo
 \else \expandafter \@secondoftwo
 \fi
}%
\providecommand \natexlab [1]{#1}%
\providecommand \enquote  [1]{``#1''}%
\providecommand \bibnamefont  [1]{#1}%
\providecommand \bibfnamefont [1]{#1}%
\providecommand \citenamefont [1]{#1}%
\providecommand \href@noop [0]{\@secondoftwo}%
\providecommand \href [0]{\begingroup \@sanitize@url \@href}%
\providecommand \@href[1]{\@@startlink{#1}\@@href}%
\providecommand \@@href[1]{\endgroup#1\@@endlink}%
\providecommand \@sanitize@url [0]{\catcode `\\12\catcode `\$12\catcode
  `\&12\catcode `\#12\catcode `\^12\catcode `\_12\catcode `\%12\relax}%
\providecommand \@@startlink[1]{}%
\providecommand \@@endlink[0]{}%
\providecommand \url  [0]{\begingroup\@sanitize@url \@url }%
\providecommand \@url [1]{\endgroup\@href {#1}{\urlprefix }}%
\providecommand \urlprefix  [0]{URL }%
\providecommand \Eprint [0]{\href }%
\providecommand \doibase [0]{http://dx.doi.org/}%
\providecommand \selectlanguage [0]{\@gobble}%
\providecommand \bibinfo  [0]{\@secondoftwo}%
\providecommand \bibfield  [0]{\@secondoftwo}%
\providecommand \translation [1]{[#1]}%
\providecommand \BibitemOpen [0]{}%
\providecommand \bibitemStop [0]{}%
\providecommand \bibitemNoStop [0]{.\EOS\space}%
\providecommand \EOS [0]{\spacefactor3000\relax}%
\providecommand \BibitemShut  [1]{\csname bibitem#1\endcsname}%
\let\auto@bib@innerbib\@empty
\bibitem [{\citenamefont {Tanabashi}\ \emph {et~al.}(2018)\citenamefont
  {Tanabashi} \emph {et~al.}}]{PhysRevD.98.030001}%
  \BibitemOpen
  \bibfield  {author} {\bibinfo {author} {\bibfnamefont {M.}~\bibnamefont
  {Tanabashi}} \emph {et~al.} (\bibinfo {collaboration} {Particle Data
  Group}),\ }\href {\doibase 10.1103/PhysRevD.98.030001} {\bibfield  {journal}
  {\bibinfo  {journal} {Phys. Rev. D}\ }\textbf {\bibinfo {volume} {98}},\
  \bibinfo {pages} {030001} (\bibinfo {year} {2018})}\BibitemShut {NoStop}%
\bibitem [{\citenamefont {Liu}\ and\ \citenamefont {Dong}(1994)}]{Liu:1993cv}%
  \BibitemOpen
  \bibfield  {author} {\bibinfo {author} {\bibfnamefont {K.-F.}\ \bibnamefont
  {Liu}}\ and\ \bibinfo {author} {\bibfnamefont {S.-J.}\ \bibnamefont {Dong}},\
  }\href {\doibase 10.1103/PhysRevLett.72.1790} {\bibfield  {journal} {\bibinfo
   {journal} {Phys. Rev. Lett.}\ }\textbf {\bibinfo {volume} {72}},\ \bibinfo
  {pages} {1790} (\bibinfo {year} {1994})},\ \Eprint
  {http://arxiv.org/abs/hep-ph/9306299} {arXiv:hep-ph/9306299 [hep-ph]}
  \BibitemShut {NoStop}%
\bibitem [{\citenamefont {Aglietti}\ \emph {et~al.}(1998)\citenamefont
  {Aglietti}, \citenamefont {Ciuchini}, \citenamefont {Corbo}, \citenamefont
  {Franco}, \citenamefont {Martinelli},\ and\ \citenamefont
  {Silvestrini}}]{Aglietti:1998mz}%
  \BibitemOpen
  \bibfield  {author} {\bibinfo {author} {\bibfnamefont {U.}~\bibnamefont
  {Aglietti}}, \bibinfo {author} {\bibfnamefont {M.}~\bibnamefont {Ciuchini}},
  \bibinfo {author} {\bibfnamefont {G.}~\bibnamefont {Corbo}}, \bibinfo
  {author} {\bibfnamefont {E.}~\bibnamefont {Franco}}, \bibinfo {author}
  {\bibfnamefont {G.}~\bibnamefont {Martinelli}}, \ and\ \bibinfo {author}
  {\bibfnamefont {L.}~\bibnamefont {Silvestrini}},\ }\href {\doibase
  10.1016/S0370-2693(98)00677-7} {\bibfield  {journal} {\bibinfo  {journal}
  {Phys. Lett.}\ }\textbf {\bibinfo {volume} {B432}},\ \bibinfo {pages} {411}
  (\bibinfo {year} {1998})},\ \Eprint {http://arxiv.org/abs/hep-ph/9804416}
  {arXiv:hep-ph/9804416 [hep-ph]} \BibitemShut {NoStop}%
\bibitem [{\citenamefont {Liu}(2000)}]{Liu:1999ak}%
  \BibitemOpen
  \bibfield  {author} {\bibinfo {author} {\bibfnamefont {K.-F.}\ \bibnamefont
  {Liu}},\ }\href {\doibase 10.1103/PhysRevD.62.074501} {\bibfield  {journal}
  {\bibinfo  {journal} {Phys. Rev.}\ }\textbf {\bibinfo {volume} {D62}},\
  \bibinfo {pages} {074501} (\bibinfo {year} {2000})},\ \Eprint
  {http://arxiv.org/abs/hep-ph/9910306} {arXiv:hep-ph/9910306 [hep-ph]}
  \BibitemShut {NoStop}%
\bibitem [{\citenamefont {Ji}(2013)}]{Ji:2013dva}%
  \BibitemOpen
  \bibfield  {author} {\bibinfo {author} {\bibfnamefont {X.}~\bibnamefont
  {Ji}},\ }\href {\doibase 10.1103/PhysRevLett.110.262002} {\bibfield
  {journal} {\bibinfo  {journal} {Phys. Rev. Lett.}\ }\textbf {\bibinfo
  {volume} {110}},\ \bibinfo {pages} {262002} (\bibinfo {year} {2013})},\
  \Eprint {http://arxiv.org/abs/1305.1539} {arXiv:1305.1539 [hep-ph]}
  \BibitemShut {NoStop}%
\bibitem [{\citenamefont {Chambers}\ \emph {et~al.}(2017)\citenamefont
  {Chambers}, \citenamefont {Horsley}, \citenamefont {Nakamura}, \citenamefont
  {Perlt}, \citenamefont {Rakow}, \citenamefont {Schierholz}, \citenamefont
  {Schiller}, \citenamefont {Somfleth}, \citenamefont {Young},\ and\
  \citenamefont {Zanotti}}]{Chambers:2017dov}%
  \BibitemOpen
  \bibfield  {author} {\bibinfo {author} {\bibfnamefont {A.~J.}\ \bibnamefont
  {Chambers}}, \bibinfo {author} {\bibfnamefont {R.}~\bibnamefont {Horsley}},
  \bibinfo {author} {\bibfnamefont {Y.}~\bibnamefont {Nakamura}}, \bibinfo
  {author} {\bibfnamefont {H.}~\bibnamefont {Perlt}}, \bibinfo {author}
  {\bibfnamefont {P.~E.~L.}\ \bibnamefont {Rakow}}, \bibinfo {author}
  {\bibfnamefont {G.}~\bibnamefont {Schierholz}}, \bibinfo {author}
  {\bibfnamefont {A.}~\bibnamefont {Schiller}}, \bibinfo {author}
  {\bibfnamefont {K.}~\bibnamefont {Somfleth}}, \bibinfo {author}
  {\bibfnamefont {R.~D.}\ \bibnamefont {Young}}, \ and\ \bibinfo {author}
  {\bibfnamefont {J.~M.}\ \bibnamefont {Zanotti}},\ }\href {\doibase
  10.1103/PhysRevLett.118.242001} {\bibfield  {journal} {\bibinfo  {journal}
  {Phys. Rev. Lett.}\ }\textbf {\bibinfo {volume} {118}},\ \bibinfo {pages}
  {242001} (\bibinfo {year} {2017})},\ \Eprint
  {http://arxiv.org/abs/1703.01153} {arXiv:1703.01153 [hep-lat]} \BibitemShut
  {NoStop}%
\bibitem [{\citenamefont {Bali}\ \emph {et~al.}(2018)\citenamefont {Bali},
  \citenamefont {Braun}, \citenamefont {Gl{\"a\ss}le}, \citenamefont
  {G{\"o}ckeler}, \citenamefont {Gruber}, \citenamefont {Hutzler},
  \citenamefont {Korcyl}, \citenamefont {Sch{\"af}er}, \citenamefont {Wein},\
  and\ \citenamefont {Zhang}}]{Bali:2018spj}%
  \BibitemOpen
  \bibfield  {author} {\bibinfo {author} {\bibfnamefont {G.~S.}\ \bibnamefont
  {Bali}}, \bibinfo {author} {\bibfnamefont {V.~M.}\ \bibnamefont {Braun}},
  \bibinfo {author} {\bibfnamefont {B.}~\bibnamefont {Gl{\"a\ss}le}}, \bibinfo
  {author} {\bibfnamefont {M.}~\bibnamefont {G{\"o}ckeler}}, \bibinfo {author}
  {\bibfnamefont {M.}~\bibnamefont {Gruber}}, \bibinfo {author} {\bibfnamefont
  {F.}~\bibnamefont {Hutzler}}, \bibinfo {author} {\bibfnamefont
  {P.}~\bibnamefont {Korcyl}}, \bibinfo {author} {\bibfnamefont
  {A.}~\bibnamefont {Sch{\"af}er}}, \bibinfo {author} {\bibfnamefont
  {P.}~\bibnamefont {Wein}}, \ and\ \bibinfo {author} {\bibfnamefont {J.-H.}\
  \bibnamefont {Zhang}},\ }\href {\doibase 10.1103/PhysRevD.98.094507}
  {\bibfield  {journal} {\bibinfo  {journal} {Phys. Rev.}\ }\textbf {\bibinfo
  {volume} {D98}},\ \bibinfo {pages} {094507} (\bibinfo {year} {2018})},\
  \Eprint {http://arxiv.org/abs/1807.06671} {arXiv:1807.06671 [hep-lat]}
  \BibitemShut {NoStop}%
\bibitem [{\citenamefont {Radyushkin}(2017)}]{Radyushkin:2017cyf}%
  \BibitemOpen
  \bibfield  {author} {\bibinfo {author} {\bibfnamefont {A.~V.}\ \bibnamefont
  {Radyushkin}},\ }\href {\doibase 10.1103/PhysRevD.96.034025} {\bibfield
  {journal} {\bibinfo  {journal} {Phys. Rev.}\ }\textbf {\bibinfo {volume}
  {D96}},\ \bibinfo {pages} {034025} (\bibinfo {year} {2017})},\ \Eprint
  {http://arxiv.org/abs/1705.01488} {arXiv:1705.01488 [hep-ph]} \BibitemShut
  {NoStop}%
\bibitem [{\citenamefont {Orginos}\ \emph {et~al.}(2017)\citenamefont
  {Orginos}, \citenamefont {Radyushkin}, \citenamefont {Karpie},\ and\
  \citenamefont {Zafeiropoulos}}]{Orginos:2017kos}%
  \BibitemOpen
  \bibfield  {author} {\bibinfo {author} {\bibfnamefont {K.}~\bibnamefont
  {Orginos}}, \bibinfo {author} {\bibfnamefont {A.}~\bibnamefont {Radyushkin}},
  \bibinfo {author} {\bibfnamefont {J.}~\bibnamefont {Karpie}}, \ and\ \bibinfo
  {author} {\bibfnamefont {S.}~\bibnamefont {Zafeiropoulos}},\ }\href {\doibase
  10.1103/PhysRevD.96.094503} {\bibfield  {journal} {\bibinfo  {journal} {Phys.
  Rev.}\ }\textbf {\bibinfo {volume} {D96}},\ \bibinfo {pages} {094503}
  (\bibinfo {year} {2017})},\ \Eprint {http://arxiv.org/abs/1706.05373}
  {arXiv:1706.05373 [hep-ph]} \BibitemShut {NoStop}%
\bibitem [{\citenamefont {Karpie}\ \emph {et~al.}(2018)\citenamefont {Karpie},
  \citenamefont {Orginos},\ and\ \citenamefont
  {Zafeiropoulos}}]{Karpie:2018zaz}%
  \BibitemOpen
  \bibfield  {author} {\bibinfo {author} {\bibfnamefont {J.}~\bibnamefont
  {Karpie}}, \bibinfo {author} {\bibfnamefont {K.}~\bibnamefont {Orginos}}, \
  and\ \bibinfo {author} {\bibfnamefont {S.}~\bibnamefont {Zafeiropoulos}},\
  }\href {\doibase 10.1007/JHEP11(2018)178} {\bibfield  {journal} {\bibinfo
  {journal} {JHEP}\ }\textbf {\bibinfo {volume} {11}},\ \bibinfo {pages} {178}
  (\bibinfo {year} {2018})},\ \Eprint {http://arxiv.org/abs/1807.10933}
  {arXiv:1807.10933 [hep-lat]} \BibitemShut {NoStop}%
\bibitem [{\citenamefont {Karpie}\ \emph {et~al.}(2019)\citenamefont {Karpie},
  \citenamefont {Orginos}, \citenamefont {Rothkopf},\ and\ \citenamefont
  {Zafeiropoulos}}]{Karpie:2019eiq}%
  \BibitemOpen
  \bibfield  {author} {\bibinfo {author} {\bibfnamefont {J.}~\bibnamefont
  {Karpie}}, \bibinfo {author} {\bibfnamefont {K.}~\bibnamefont {Orginos}},
  \bibinfo {author} {\bibfnamefont {A.}~\bibnamefont {Rothkopf}}, \ and\
  \bibinfo {author} {\bibfnamefont {S.}~\bibnamefont {Zafeiropoulos}},\ }\href
  {\doibase 10.1007/JHEP04(2019)057} {\bibfield  {journal} {\bibinfo  {journal}
  {JHEP}\ }\textbf {\bibinfo {volume} {04}},\ \bibinfo {pages} {057} (\bibinfo
  {year} {2019})},\ \Eprint {http://arxiv.org/abs/1901.05408} {arXiv:1901.05408
  [hep-lat]} \BibitemShut {NoStop}%
\bibitem [{\citenamefont {Cichy}\ \emph {et~al.}(2019)\citenamefont {Cichy},
  \citenamefont {Del~Debbio},\ and\ \citenamefont {Giani}}]{Cichy:2019ebf}%
  \BibitemOpen
  \bibfield  {author} {\bibinfo {author} {\bibfnamefont {K.}~\bibnamefont
  {Cichy}}, \bibinfo {author} {\bibfnamefont {L.}~\bibnamefont {Del~Debbio}}, \
  and\ \bibinfo {author} {\bibfnamefont {T.}~\bibnamefont {Giani}},\
  }\href@noop {} {\  (\bibinfo {year} {2019})},\ \Eprint
  {http://arxiv.org/abs/1907.06037} {arXiv:1907.06037 [hep-ph]} \BibitemShut
  {NoStop}%
\bibitem [{\citenamefont {Alexandrou}\ \emph {et~al.}(2019)\citenamefont
  {Alexandrou}, \citenamefont {Cichy}, \citenamefont {Constantinou},
  \citenamefont {Hadjiyiannakou}, \citenamefont {Jansen}, \citenamefont
  {Scapellato},\ and\ \citenamefont {Steffens}}]{Alexandrou:2019lfo}%
  \BibitemOpen
  \bibfield  {author} {\bibinfo {author} {\bibfnamefont {C.}~\bibnamefont
  {Alexandrou}}, \bibinfo {author} {\bibfnamefont {K.}~\bibnamefont {Cichy}},
  \bibinfo {author} {\bibfnamefont {M.}~\bibnamefont {Constantinou}}, \bibinfo
  {author} {\bibfnamefont {K.}~\bibnamefont {Hadjiyiannakou}}, \bibinfo
  {author} {\bibfnamefont {K.}~\bibnamefont {Jansen}}, \bibinfo {author}
  {\bibfnamefont {A.}~\bibnamefont {Scapellato}}, \ and\ \bibinfo {author}
  {\bibfnamefont {F.}~\bibnamefont {Steffens}},\ }\href {\doibase
  10.1103/PhysRevD.99.114504} {\bibfield  {journal} {\bibinfo  {journal} {Phys.
  Rev.}\ }\textbf {\bibinfo {volume} {D99}},\ \bibinfo {pages} {114504}
  (\bibinfo {year} {2019})},\ \Eprint {http://arxiv.org/abs/1902.00587}
  {arXiv:1902.00587 [hep-lat]} \BibitemShut {NoStop}%
\bibitem [{\citenamefont {Sufian}\ \emph {et~al.}(2019)\citenamefont {Sufian},
  \citenamefont {Karpie}, \citenamefont {Egerer}, \citenamefont {Orginos},
  \citenamefont {Qiu},\ and\ \citenamefont {Richards}}]{Sufian:2019bol}%
  \BibitemOpen
  \bibfield  {author} {\bibinfo {author} {\bibfnamefont {R.~S.}\ \bibnamefont
  {Sufian}}, \bibinfo {author} {\bibfnamefont {J.}~\bibnamefont {Karpie}},
  \bibinfo {author} {\bibfnamefont {C.}~\bibnamefont {Egerer}}, \bibinfo
  {author} {\bibfnamefont {K.}~\bibnamefont {Orginos}}, \bibinfo {author}
  {\bibfnamefont {J.-W.}\ \bibnamefont {Qiu}}, \ and\ \bibinfo {author}
  {\bibfnamefont {D.~G.}\ \bibnamefont {Richards}},\ }\href {\doibase
  10.1103/PhysRevD.99.074507} {\bibfield  {journal} {\bibinfo  {journal} {Phys.
  Rev.}\ }\textbf {\bibinfo {volume} {D99}},\ \bibinfo {pages} {074507}
  (\bibinfo {year} {2019})},\ \Eprint {http://arxiv.org/abs/1901.03921}
  {arXiv:1901.03921 [hep-lat]} \BibitemShut {NoStop}%
\bibitem [{\citenamefont {Ma}\ and\ \citenamefont
  {Qiu}(2018{\natexlab{a}})}]{Ma:2014jla}%
  \BibitemOpen
  \bibfield  {author} {\bibinfo {author} {\bibfnamefont {Y.-Q.}\ \bibnamefont
  {Ma}}\ and\ \bibinfo {author} {\bibfnamefont {J.-W.}\ \bibnamefont {Qiu}},\
  }\href {\doibase 10.1103/PhysRevD.98.074021} {\bibfield  {journal} {\bibinfo
  {journal} {Phys. Rev.}\ }\textbf {\bibinfo {volume} {D98}},\ \bibinfo {pages}
  {074021} (\bibinfo {year} {2018}{\natexlab{a}})},\ \Eprint
  {http://arxiv.org/abs/1404.6860} {arXiv:1404.6860 [hep-ph]} \BibitemShut
  {NoStop}%
\bibitem [{\citenamefont {Ma}\ and\ \citenamefont {Qiu}(2015)}]{Ma:2014jga}%
  \BibitemOpen
  \bibfield  {author} {\bibinfo {author} {\bibfnamefont {Y.-Q.}\ \bibnamefont
  {Ma}}\ and\ \bibinfo {author} {\bibfnamefont {J.-W.}\ \bibnamefont {Qiu}},\
  }\bibfield  {booktitle} {\emph {\bibinfo {booktitle} {{Proceedings, QCD
  Evolution Workshop (QCD 2014): Santa Fe, USA, May 12-16, 2014}}},\ }\href
  {\doibase 10.1142/S2010194515600411} {\bibfield  {journal} {\bibinfo
  {journal} {Int. J. Mod. Phys. Conf. Ser.}\ }\textbf {\bibinfo {volume}
  {37}},\ \bibinfo {pages} {1560041} (\bibinfo {year} {2015})},\ \Eprint
  {http://arxiv.org/abs/1412.2688} {arXiv:1412.2688 [hep-ph]} \BibitemShut
  {NoStop}%
\bibitem [{\citenamefont {Ma}\ and\ \citenamefont
  {Qiu}(2018{\natexlab{b}})}]{Ma:2017pxb}%
  \BibitemOpen
  \bibfield  {author} {\bibinfo {author} {\bibfnamefont {Y.-Q.}\ \bibnamefont
  {Ma}}\ and\ \bibinfo {author} {\bibfnamefont {J.-W.}\ \bibnamefont {Qiu}},\
  }\href {\doibase 10.1103/PhysRevLett.120.022003} {\bibfield  {journal}
  {\bibinfo  {journal} {Phys. Rev. Lett.}\ }\textbf {\bibinfo {volume} {120}},\
  \bibinfo {pages} {022003} (\bibinfo {year} {2018}{\natexlab{b}})},\ \Eprint
  {http://arxiv.org/abs/1709.03018} {arXiv:1709.03018 [hep-ph]} \BibitemShut
  {NoStop}%
\bibitem [{\citenamefont {Liang}\ \emph {et~al.}(2019)\citenamefont {Liang},
  \citenamefont {Draper}, \citenamefont {Liu}, \citenamefont {Rothkopf},\ and\
  \citenamefont {Yang}}]{Liang:2019frk}%
  \BibitemOpen
  \bibfield  {author} {\bibinfo {author} {\bibfnamefont {J.}~\bibnamefont
  {Liang}}, \bibinfo {author} {\bibfnamefont {T.}~\bibnamefont {Draper}},
  \bibinfo {author} {\bibfnamefont {K.-F.}\ \bibnamefont {Liu}}, \bibinfo
  {author} {\bibfnamefont {A.}~\bibnamefont {Rothkopf}}, \ and\ \bibinfo
  {author} {\bibfnamefont {Y.-B.}\ \bibnamefont {Yang}} (\bibinfo
  {collaboration} {XQCD}),\ }\href@noop {} {\  (\bibinfo {year} {2019})},\
  \Eprint {http://arxiv.org/abs/1906.05312} {arXiv:1906.05312 [hep-ph]}
  \BibitemShut {NoStop}%
\bibitem [{\citenamefont {Banerjee}\ \emph {et~al.}(2012)\citenamefont
  {Banerjee}, \citenamefont {Dalmonte}, \citenamefont {Muller}, \citenamefont
  {Rico}, \citenamefont {Stebler}, \citenamefont {Wiese},\ and\ \citenamefont
  {Zoller}}]{Banerjee:2012pg}%
  \BibitemOpen
  \bibfield  {author} {\bibinfo {author} {\bibfnamefont {D.}~\bibnamefont
  {Banerjee}}, \bibinfo {author} {\bibfnamefont {M.}~\bibnamefont {Dalmonte}},
  \bibinfo {author} {\bibfnamefont {M.}~\bibnamefont {Muller}}, \bibinfo
  {author} {\bibfnamefont {E.}~\bibnamefont {Rico}}, \bibinfo {author}
  {\bibfnamefont {P.}~\bibnamefont {Stebler}}, \bibinfo {author} {\bibfnamefont
  {U.~J.}\ \bibnamefont {Wiese}}, \ and\ \bibinfo {author} {\bibfnamefont
  {P.}~\bibnamefont {Zoller}},\ }\href {\doibase
  10.1103/PhysRevLett.109.175302} {\bibfield  {journal} {\bibinfo  {journal}
  {Phys. Rev. Lett.}\ }\textbf {\bibinfo {volume} {109}},\ \bibinfo {pages}
  {175302} (\bibinfo {year} {2012})},\ \Eprint {http://arxiv.org/abs/1205.6366}
  {arXiv:1205.6366 [cond-mat.quant-gas]} \BibitemShut {NoStop}%
\bibitem [{\citenamefont {Hauke}\ \emph {et~al.}(2013)\citenamefont {Hauke},
  \citenamefont {Marcos}, \citenamefont {Dalmonte},\ and\ \citenamefont
  {Zoller}}]{Hauke:2013jga}%
  \BibitemOpen
  \bibfield  {author} {\bibinfo {author} {\bibfnamefont {P.}~\bibnamefont
  {Hauke}}, \bibinfo {author} {\bibfnamefont {D.}~\bibnamefont {Marcos}},
  \bibinfo {author} {\bibfnamefont {M.}~\bibnamefont {Dalmonte}}, \ and\
  \bibinfo {author} {\bibfnamefont {P.}~\bibnamefont {Zoller}},\ }\href
  {\doibase 10.1103/PhysRevX.3.041018} {\bibfield  {journal} {\bibinfo
  {journal} {Phys. Rev.}\ }\textbf {\bibinfo {volume} {X3}},\ \bibinfo {pages}
  {041018} (\bibinfo {year} {2013})},\ \Eprint {http://arxiv.org/abs/1306.2162}
  {arXiv:1306.2162 [cond-mat.quant-gas]} \BibitemShut {NoStop}%
\bibitem [{\citenamefont {Kuno}\ \emph {et~al.}(2015)\citenamefont {Kuno},
  \citenamefont {Kasamatsu}, \citenamefont {Takahashi}, \citenamefont
  {Ichinose},\ and\ \citenamefont {Matsui}}]{Kuno:2014npa}%
  \BibitemOpen
  \bibfield  {author} {\bibinfo {author} {\bibfnamefont {Y.}~\bibnamefont
  {Kuno}}, \bibinfo {author} {\bibfnamefont {K.}~\bibnamefont {Kasamatsu}},
  \bibinfo {author} {\bibfnamefont {Y.}~\bibnamefont {Takahashi}}, \bibinfo
  {author} {\bibfnamefont {I.}~\bibnamefont {Ichinose}}, \ and\ \bibinfo
  {author} {\bibfnamefont {T.}~\bibnamefont {Matsui}},\ }\href {\doibase
  10.1088/1367-2630/17/6/063005} {\bibfield  {journal} {\bibinfo  {journal}
  {New J. Phys.}\ }\textbf {\bibinfo {volume} {17}},\ \bibinfo {pages} {063005}
  (\bibinfo {year} {2015})},\ \Eprint {http://arxiv.org/abs/1412.7605}
  {arXiv:1412.7605 [cond-mat.quant-gas]} \BibitemShut {NoStop}%
\bibitem [{\citenamefont {Kuno}\ \emph {et~al.}(2017)\citenamefont {Kuno},
  \citenamefont {Sakane}, \citenamefont {Kasamatsu}, \citenamefont {Ichinose},\
  and\ \citenamefont {Matsui}}]{Kuno:2016ipi}%
  \BibitemOpen
  \bibfield  {author} {\bibinfo {author} {\bibfnamefont {Y.}~\bibnamefont
  {Kuno}}, \bibinfo {author} {\bibfnamefont {S.}~\bibnamefont {Sakane}},
  \bibinfo {author} {\bibfnamefont {K.}~\bibnamefont {Kasamatsu}}, \bibinfo
  {author} {\bibfnamefont {I.}~\bibnamefont {Ichinose}}, \ and\ \bibinfo
  {author} {\bibfnamefont {T.}~\bibnamefont {Matsui}},\ }\href {\doibase
  10.1103/PhysRevD.95.094507} {\bibfield  {journal} {\bibinfo  {journal} {Phys.
  Rev.}\ }\textbf {\bibinfo {volume} {D95}},\ \bibinfo {pages} {094507}
  (\bibinfo {year} {2017})},\ \Eprint {http://arxiv.org/abs/1605.00333}
  {arXiv:1605.00333 [cond-mat.quant-gas]} \BibitemShut {NoStop}%
\bibitem [{\citenamefont {Martinez}\ \emph {et~al.}(2016)\citenamefont
  {Martinez} \emph {et~al.}}]{Martinez:2016yna}%
  \BibitemOpen
  \bibfield  {author} {\bibinfo {author} {\bibfnamefont {E.~A.}\ \bibnamefont
  {Martinez}} \emph {et~al.},\ }\href {\doibase 10.1038/nature18318} {\bibfield
   {journal} {\bibinfo  {journal} {Nature}\ }\textbf {\bibinfo {volume}
  {534}},\ \bibinfo {pages} {516} (\bibinfo {year} {2016})},\ \Eprint
  {http://arxiv.org/abs/1605.04570} {arXiv:1605.04570 [quant-ph]} \BibitemShut
  {NoStop}%
\bibitem [{\citenamefont {Marcos}\ \emph {et~al.}(2014)\citenamefont {Marcos},
  \citenamefont {Widmer}, \citenamefont {Rico}, \citenamefont {Hafezi},
  \citenamefont {Rabl}, \citenamefont {Wiese},\ and\ \citenamefont
  {Zoller}}]{Marcos:2014lda}%
  \BibitemOpen
  \bibfield  {author} {\bibinfo {author} {\bibfnamefont {D.}~\bibnamefont
  {Marcos}}, \bibinfo {author} {\bibfnamefont {P.}~\bibnamefont {Widmer}},
  \bibinfo {author} {\bibfnamefont {E.}~\bibnamefont {Rico}}, \bibinfo {author}
  {\bibfnamefont {M.}~\bibnamefont {Hafezi}}, \bibinfo {author} {\bibfnamefont
  {P.}~\bibnamefont {Rabl}}, \bibinfo {author} {\bibfnamefont {U.~J.}\
  \bibnamefont {Wiese}}, \ and\ \bibinfo {author} {\bibfnamefont
  {P.}~\bibnamefont {Zoller}},\ }\href {\doibase 10.1016/j.aop.2014.09.011}
  {\bibfield  {journal} {\bibinfo  {journal} {Annals Phys.}\ }\textbf {\bibinfo
  {volume} {351}},\ \bibinfo {pages} {634} (\bibinfo {year} {2014})},\ \Eprint
  {http://arxiv.org/abs/1407.6066} {arXiv:1407.6066 [quant-ph]} \BibitemShut
  {NoStop}%
\bibitem [{\citenamefont {Klco}\ \emph {et~al.}(2018)\citenamefont {Klco},
  \citenamefont {Dumitrescu}, \citenamefont {McCaskey}, \citenamefont {Morris},
  \citenamefont {Pooser}, \citenamefont {Sanz}, \citenamefont {Solano},
  \citenamefont {Lougovski},\ and\ \citenamefont {Savage}}]{Klco:2018kyo}%
  \BibitemOpen
  \bibfield  {author} {\bibinfo {author} {\bibfnamefont {N.}~\bibnamefont
  {Klco}}, \bibinfo {author} {\bibfnamefont {E.~F.}\ \bibnamefont
  {Dumitrescu}}, \bibinfo {author} {\bibfnamefont {A.~J.}\ \bibnamefont
  {McCaskey}}, \bibinfo {author} {\bibfnamefont {T.~D.}\ \bibnamefont
  {Morris}}, \bibinfo {author} {\bibfnamefont {R.~C.}\ \bibnamefont {Pooser}},
  \bibinfo {author} {\bibfnamefont {M.}~\bibnamefont {Sanz}}, \bibinfo {author}
  {\bibfnamefont {E.}~\bibnamefont {Solano}}, \bibinfo {author} {\bibfnamefont
  {P.}~\bibnamefont {Lougovski}}, \ and\ \bibinfo {author} {\bibfnamefont
  {M.~J.}\ \bibnamefont {Savage}},\ }\href {\doibase
  10.1103/PhysRevA.98.032331} {\bibfield  {journal} {\bibinfo  {journal} {Phys.
  Rev.}\ }\textbf {\bibinfo {volume} {A98}},\ \bibinfo {pages} {032331}
  (\bibinfo {year} {2018})},\ \Eprint {http://arxiv.org/abs/1803.03326}
  {arXiv:1803.03326 [quant-ph]} \BibitemShut {NoStop}%
\bibitem [{\citenamefont {Zache}\ \emph {et~al.}(2018)\citenamefont {Zache},
  \citenamefont {Hebenstreit}, \citenamefont {Jendrzejewski}, \citenamefont
  {Oberthaler}, \citenamefont {Berges},\ and\ \citenamefont
  {Hauke}}]{Zache:2018jbt}%
  \BibitemOpen
  \bibfield  {author} {\bibinfo {author} {\bibfnamefont {T.~V.}\ \bibnamefont
  {Zache}}, \bibinfo {author} {\bibfnamefont {F.}~\bibnamefont {Hebenstreit}},
  \bibinfo {author} {\bibfnamefont {F.}~\bibnamefont {Jendrzejewski}}, \bibinfo
  {author} {\bibfnamefont {M.~K.}\ \bibnamefont {Oberthaler}}, \bibinfo
  {author} {\bibfnamefont {J.}~\bibnamefont {Berges}}, \ and\ \bibinfo {author}
  {\bibfnamefont {P.}~\bibnamefont {Hauke}},\ }\href {\doibase
  10.1088/2058-9565/aac33b} {\bibfield  {journal} {\bibinfo  {journal} {Sci.
  Technol.}\ }\textbf {\bibinfo {volume} {3}},\ \bibinfo {pages} {034010}
  (\bibinfo {year} {2018})},\ \Eprint {http://arxiv.org/abs/1802.06704}
  {arXiv:1802.06704 [cond-mat.quant-gas]} \BibitemShut {NoStop}%
\bibitem [{\citenamefont {Muschik}\ \emph {et~al.}(2017)\citenamefont
  {Muschik}, \citenamefont {Heyl}, \citenamefont {Martinez}, \citenamefont
  {Monz}, \citenamefont {Schindler}, \citenamefont {Vogell}, \citenamefont
  {Dalmonte}, \citenamefont {Hauke}, \citenamefont {Blatt},\ and\ \citenamefont
  {Zoller}}]{Muschik:2016tws}%
  \BibitemOpen
  \bibfield  {author} {\bibinfo {author} {\bibfnamefont {C.}~\bibnamefont
  {Muschik}}, \bibinfo {author} {\bibfnamefont {M.}~\bibnamefont {Heyl}},
  \bibinfo {author} {\bibfnamefont {E.}~\bibnamefont {Martinez}}, \bibinfo
  {author} {\bibfnamefont {T.}~\bibnamefont {Monz}}, \bibinfo {author}
  {\bibfnamefont {P.}~\bibnamefont {Schindler}}, \bibinfo {author}
  {\bibfnamefont {B.}~\bibnamefont {Vogell}}, \bibinfo {author} {\bibfnamefont
  {M.}~\bibnamefont {Dalmonte}}, \bibinfo {author} {\bibfnamefont
  {P.}~\bibnamefont {Hauke}}, \bibinfo {author} {\bibfnamefont
  {R.}~\bibnamefont {Blatt}}, \ and\ \bibinfo {author} {\bibfnamefont
  {P.}~\bibnamefont {Zoller}},\ }\href {\doibase 10.1088/1367-2630/aa89ab}
  {\bibfield  {journal} {\bibinfo  {journal} {New J. Phys.}\ }\textbf {\bibinfo
  {volume} {19}},\ \bibinfo {pages} {103020} (\bibinfo {year} {2017})},\
  \Eprint {http://arxiv.org/abs/1612.08653} {arXiv:1612.08653 [quant-ph]}
  \BibitemShut {NoStop}%
\bibitem [{\citenamefont {Bazavov}\ \emph {et~al.}(2015)\citenamefont
  {Bazavov}, \citenamefont {Meurice}, \citenamefont {Tsai}, \citenamefont
  {Unmuth-Yockey},\ and\ \citenamefont {Zhang}}]{Bazavov:2015kka}%
  \BibitemOpen
  \bibfield  {author} {\bibinfo {author} {\bibfnamefont {A.}~\bibnamefont
  {Bazavov}}, \bibinfo {author} {\bibfnamefont {Y.}~\bibnamefont {Meurice}},
  \bibinfo {author} {\bibfnamefont {S.-W.}\ \bibnamefont {Tsai}}, \bibinfo
  {author} {\bibfnamefont {J.}~\bibnamefont {Unmuth-Yockey}}, \ and\ \bibinfo
  {author} {\bibfnamefont {J.}~\bibnamefont {Zhang}},\ }\href {\doibase
  10.1103/PhysRevD.92.076003} {\bibfield  {journal} {\bibinfo  {journal} {Phys.
  Rev.}\ }\textbf {\bibinfo {volume} {D92}},\ \bibinfo {pages} {076003}
  (\bibinfo {year} {2015})},\ \Eprint {http://arxiv.org/abs/1503.08354}
  {arXiv:1503.08354 [hep-lat]} \BibitemShut {NoStop}%
\bibitem [{\citenamefont {Zhang}\ \emph {et~al.}(2018)\citenamefont {Zhang},
  \citenamefont {Unmuth-Yockey}, \citenamefont {Zeiher}, \citenamefont
  {Bazavov}, \citenamefont {Tsai},\ and\ \citenamefont
  {Meurice}}]{Zhang:2018ufj}%
  \BibitemOpen
  \bibfield  {author} {\bibinfo {author} {\bibfnamefont {J.}~\bibnamefont
  {Zhang}}, \bibinfo {author} {\bibfnamefont {J.}~\bibnamefont
  {Unmuth-Yockey}}, \bibinfo {author} {\bibfnamefont {J.}~\bibnamefont
  {Zeiher}}, \bibinfo {author} {\bibfnamefont {A.}~\bibnamefont {Bazavov}},
  \bibinfo {author} {\bibfnamefont {S.~W.}\ \bibnamefont {Tsai}}, \ and\
  \bibinfo {author} {\bibfnamefont {Y.}~\bibnamefont {Meurice}},\ }\href
  {\doibase 10.1103/PhysRevLett.121.223201} {\bibfield  {journal} {\bibinfo
  {journal} {Phys. Rev. Lett.}\ }\textbf {\bibinfo {volume} {121}},\ \bibinfo
  {pages} {223201} (\bibinfo {year} {2018})},\ \Eprint
  {http://arxiv.org/abs/1803.11166} {arXiv:1803.11166 [hep-lat]} \BibitemShut
  {NoStop}%
\bibitem [{\citenamefont {Unmuth-Yockey}(2019)}]{Unmuth-Yockey:2018xak}%
  \BibitemOpen
  \bibfield  {author} {\bibinfo {author} {\bibfnamefont {J.~F.}\ \bibnamefont
  {Unmuth-Yockey}},\ }\href {\doibase 10.1103/PhysRevD.99.074502} {\bibfield
  {journal} {\bibinfo  {journal} {Phys. Rev.}\ }\textbf {\bibinfo {volume}
  {D99}},\ \bibinfo {pages} {074502} (\bibinfo {year} {2019})},\ \Eprint
  {http://arxiv.org/abs/1811.05884} {arXiv:1811.05884 [hep-lat]} \BibitemShut
  {NoStop}%
\bibitem [{\citenamefont {Unmuth-Yockey}\ \emph {et~al.}(2018)\citenamefont
  {Unmuth-Yockey}, \citenamefont {Zhang}, \citenamefont {Bazavov},
  \citenamefont {Meurice},\ and\ \citenamefont {Tsai}}]{Unmuth-Yockey:2018ugm}%
  \BibitemOpen
  \bibfield  {author} {\bibinfo {author} {\bibfnamefont {J.}~\bibnamefont
  {Unmuth-Yockey}}, \bibinfo {author} {\bibfnamefont {J.}~\bibnamefont
  {Zhang}}, \bibinfo {author} {\bibfnamefont {A.}~\bibnamefont {Bazavov}},
  \bibinfo {author} {\bibfnamefont {Y.}~\bibnamefont {Meurice}}, \ and\
  \bibinfo {author} {\bibfnamefont {S.-W.}\ \bibnamefont {Tsai}},\ }\href
  {\doibase 10.1103/PhysRevD.98.094511} {\bibfield  {journal} {\bibinfo
  {journal} {Phys. Rev.}\ }\textbf {\bibinfo {volume} {D98}},\ \bibinfo {pages}
  {094511} (\bibinfo {year} {2018})},\ \Eprint
  {http://arxiv.org/abs/1807.09186} {arXiv:1807.09186 [hep-lat]} \BibitemShut
  {NoStop}%
\bibitem [{\citenamefont {Gustafson}\ \emph {et~al.}(2019)\citenamefont
  {Gustafson}, \citenamefont {Meurice},\ and\ \citenamefont
  {Unmuth-Yockey}}]{Gustafson:2019mpk}%
  \BibitemOpen
  \bibfield  {author} {\bibinfo {author} {\bibfnamefont {E.}~\bibnamefont
  {Gustafson}}, \bibinfo {author} {\bibfnamefont {Y.}~\bibnamefont {Meurice}},
  \ and\ \bibinfo {author} {\bibfnamefont {J.}~\bibnamefont {Unmuth-Yockey}},\
  }\href@noop {} {\  (\bibinfo {year} {2019})},\ \Eprint
  {http://arxiv.org/abs/1901.05944} {arXiv:1901.05944 [hep-lat]} \BibitemShut
  {NoStop}%
\bibitem [{\citenamefont {Zohar}\ \emph {et~al.}(2012)\citenamefont {Zohar},
  \citenamefont {Cirac},\ and\ \citenamefont {Reznik}}]{Zohar:2012ay}%
  \BibitemOpen
  \bibfield  {author} {\bibinfo {author} {\bibfnamefont {E.}~\bibnamefont
  {Zohar}}, \bibinfo {author} {\bibfnamefont {J.~I.}\ \bibnamefont {Cirac}}, \
  and\ \bibinfo {author} {\bibfnamefont {B.}~\bibnamefont {Reznik}},\ }\href
  {\doibase 10.1103/PhysRevLett.109.125302} {\bibfield  {journal} {\bibinfo
  {journal} {Phys. Rev. Lett.}\ }\textbf {\bibinfo {volume} {109}},\ \bibinfo
  {pages} {125302} (\bibinfo {year} {2012})},\ \Eprint
  {http://arxiv.org/abs/1204.6574} {arXiv:1204.6574 [quant-ph]} \BibitemShut
  {NoStop}%
\bibitem [{\citenamefont {Byrnes}\ and\ \citenamefont
  {Yamamoto}(2006)}]{Byrnes:2005qx}%
  \BibitemOpen
  \bibfield  {author} {\bibinfo {author} {\bibfnamefont {T.}~\bibnamefont
  {Byrnes}}\ and\ \bibinfo {author} {\bibfnamefont {Y.}~\bibnamefont
  {Yamamoto}},\ }\href {\doibase 10.1103/PhysRevA.73.022328} {\bibfield
  {journal} {\bibinfo  {journal} {Phys. Rev.}\ }\textbf {\bibinfo {volume}
  {A73}},\ \bibinfo {pages} {022328} (\bibinfo {year} {2006})},\ \Eprint
  {http://arxiv.org/abs/quant-ph/0510027} {arXiv:quant-ph/0510027 [quant-ph]}
  \BibitemShut {NoStop}%
\bibitem [{\citenamefont {Banerjee}\ \emph {et~al.}(2013)\citenamefont
  {Banerjee}, \citenamefont {B{\"o}gli}, \citenamefont {Dalmonte},
  \citenamefont {Rico}, \citenamefont {Stebler}, \citenamefont {Wiese},\ and\
  \citenamefont {Zoller}}]{Banerjee:2012xg}%
  \BibitemOpen
  \bibfield  {author} {\bibinfo {author} {\bibfnamefont {D.}~\bibnamefont
  {Banerjee}}, \bibinfo {author} {\bibfnamefont {M.}~\bibnamefont {B{\"o}gli}},
  \bibinfo {author} {\bibfnamefont {M.}~\bibnamefont {Dalmonte}}, \bibinfo
  {author} {\bibfnamefont {E.}~\bibnamefont {Rico}}, \bibinfo {author}
  {\bibfnamefont {P.}~\bibnamefont {Stebler}}, \bibinfo {author} {\bibfnamefont
  {U.~J.}\ \bibnamefont {Wiese}}, \ and\ \bibinfo {author} {\bibfnamefont
  {P.}~\bibnamefont {Zoller}},\ }\href {\doibase
  10.1103/PhysRevLett.110.125303} {\bibfield  {journal} {\bibinfo  {journal}
  {Phys. Rev. Lett.}\ }\textbf {\bibinfo {volume} {110}},\ \bibinfo {pages}
  {125303} (\bibinfo {year} {2013})},\ \Eprint {http://arxiv.org/abs/1211.2242}
  {arXiv:1211.2242 [cond-mat.quant-gas]} \BibitemShut {NoStop}%
\bibitem [{\citenamefont {Wiese}(2013)}]{Wiese:2013uua}%
  \BibitemOpen
  \bibfield  {author} {\bibinfo {author} {\bibfnamefont {U.-J.}\ \bibnamefont
  {Wiese}},\ }\href {\doibase 10.1002/andp.201300104} {\bibfield  {journal}
  {\bibinfo  {journal} {Annalen Phys.}\ }\textbf {\bibinfo {volume} {525}},\
  \bibinfo {pages} {777} (\bibinfo {year} {2013})},\ \Eprint
  {http://arxiv.org/abs/1305.1602} {arXiv:1305.1602 [quant-ph]} \BibitemShut
  {NoStop}%
\bibitem [{\citenamefont {Zohar}\ \emph {et~al.}(2013)\citenamefont {Zohar},
  \citenamefont {Cirac},\ and\ \citenamefont {Reznik}}]{Zohar:2012xf}%
  \BibitemOpen
  \bibfield  {author} {\bibinfo {author} {\bibfnamefont {E.}~\bibnamefont
  {Zohar}}, \bibinfo {author} {\bibfnamefont {J.~I.}\ \bibnamefont {Cirac}}, \
  and\ \bibinfo {author} {\bibfnamefont {B.}~\bibnamefont {Reznik}},\ }\href
  {\doibase 10.1103/PhysRevLett.110.125304} {\bibfield  {journal} {\bibinfo
  {journal} {Phys. Rev. Lett.}\ }\textbf {\bibinfo {volume} {110}},\ \bibinfo
  {pages} {125304} (\bibinfo {year} {2013})},\ \Eprint
  {http://arxiv.org/abs/1211.2241} {arXiv:1211.2241 [quant-ph]} \BibitemShut
  {NoStop}%
\bibitem [{\citenamefont {Tagliacozzo}\ \emph {et~al.}(2013)\citenamefont
  {Tagliacozzo}, \citenamefont {Celi}, \citenamefont {Orland},\ and\
  \citenamefont {Lewenstein}}]{Tagliacozzo:2012df}%
  \BibitemOpen
  \bibfield  {author} {\bibinfo {author} {\bibfnamefont {L.}~\bibnamefont
  {Tagliacozzo}}, \bibinfo {author} {\bibfnamefont {A.}~\bibnamefont {Celi}},
  \bibinfo {author} {\bibfnamefont {P.}~\bibnamefont {Orland}}, \ and\ \bibinfo
  {author} {\bibfnamefont {M.}~\bibnamefont {Lewenstein}},\ }\href {\doibase
  10.1038/ncomms3615} {\bibfield  {journal} {\bibinfo  {journal} {Nature
  Commun.}\ }\textbf {\bibinfo {volume} {4}},\ \bibinfo {pages} {2615}
  (\bibinfo {year} {2013})},\ \Eprint {http://arxiv.org/abs/1211.2704}
  {arXiv:1211.2704 [cond-mat.quant-gas]} \BibitemShut {NoStop}%
\bibitem [{\citenamefont {Zohar}\ \emph {et~al.}(2017)\citenamefont {Zohar},
  \citenamefont {Farace}, \citenamefont {Reznik},\ and\ \citenamefont
  {Cirac}}]{Zohar:2016iic}%
  \BibitemOpen
  \bibfield  {author} {\bibinfo {author} {\bibfnamefont {E.}~\bibnamefont
  {Zohar}}, \bibinfo {author} {\bibfnamefont {A.}~\bibnamefont {Farace}},
  \bibinfo {author} {\bibfnamefont {B.}~\bibnamefont {Reznik}}, \ and\ \bibinfo
  {author} {\bibfnamefont {J.~I.}\ \bibnamefont {Cirac}},\ }\href {\doibase
  10.1103/PhysRevA.95.023604} {\bibfield  {journal} {\bibinfo  {journal} {Phys.
  Rev.}\ }\textbf {\bibinfo {volume} {A95}},\ \bibinfo {pages} {023604}
  (\bibinfo {year} {2017})},\ \Eprint {http://arxiv.org/abs/1607.08121}
  {arXiv:1607.08121 [quant-ph]} \BibitemShut {NoStop}%
\bibitem [{\citenamefont {Bender}\ \emph {et~al.}(2018)\citenamefont {Bender},
  \citenamefont {Zohar}, \citenamefont {Farace},\ and\ \citenamefont
  {Cirac}}]{Bender:2018rdp}%
  \BibitemOpen
  \bibfield  {author} {\bibinfo {author} {\bibfnamefont {J.}~\bibnamefont
  {Bender}}, \bibinfo {author} {\bibfnamefont {E.}~\bibnamefont {Zohar}},
  \bibinfo {author} {\bibfnamefont {A.}~\bibnamefont {Farace}}, \ and\ \bibinfo
  {author} {\bibfnamefont {J.~I.}\ \bibnamefont {Cirac}},\ }\href {\doibase
  10.1088/1367-2630/aadb71} {\bibfield  {journal} {\bibinfo  {journal} {New J.
  Phys.}\ }\textbf {\bibinfo {volume} {20}},\ \bibinfo {pages} {093001}
  (\bibinfo {year} {2018})},\ \Eprint {http://arxiv.org/abs/1804.02082}
  {arXiv:1804.02082 [quant-ph]} \BibitemShut {NoStop}%
\bibitem [{\citenamefont {Lamm}\ \emph {et~al.}(2019)\citenamefont {Lamm},
  \citenamefont {Lawrence},\ and\ \citenamefont {Yamauchi}}]{Lamm:2019bik}%
  \BibitemOpen
  \bibfield  {author} {\bibinfo {author} {\bibfnamefont {H.}~\bibnamefont
  {Lamm}}, \bibinfo {author} {\bibfnamefont {S.}~\bibnamefont {Lawrence}}, \
  and\ \bibinfo {author} {\bibfnamefont {Y.}~\bibnamefont {Yamauchi}} (\bibinfo
  {collaboration} {NuQS}),\ }\href@noop {} {\  (\bibinfo {year} {2019})},\
  \Eprint {http://arxiv.org/abs/1903.08807} {arXiv:1903.08807 [hep-lat]}
  \BibitemShut {NoStop}%
\bibitem [{\citenamefont {Alexandru}\ \emph {et~al.}(2019)\citenamefont
  {Alexandru}, \citenamefont {Bedaque}, \citenamefont {Harmalkar},
  \citenamefont {Lamm}, \citenamefont {Lawrence},\ and\ \citenamefont
  {Warrington}}]{Alexandru:2019nsa}%
  \BibitemOpen
  \bibfield  {author} {\bibinfo {author} {\bibfnamefont {A.}~\bibnamefont
  {Alexandru}}, \bibinfo {author} {\bibfnamefont {P.~F.}\ \bibnamefont
  {Bedaque}}, \bibinfo {author} {\bibfnamefont {S.}~\bibnamefont {Harmalkar}},
  \bibinfo {author} {\bibfnamefont {H.}~\bibnamefont {Lamm}}, \bibinfo {author}
  {\bibfnamefont {S.}~\bibnamefont {Lawrence}}, \ and\ \bibinfo {author}
  {\bibfnamefont {N.~C.}\ \bibnamefont {Warrington}} (\bibinfo {collaboration}
  {NuQS}),\ }\href@noop {} {\  (\bibinfo {year} {2019})},\ \Eprint
  {http://arxiv.org/abs/1906.11213} {arXiv:1906.11213 [hep-lat]} \BibitemShut
  {NoStop}%
\bibitem [{\citenamefont {Klco}\ \emph {et~al.}(2019)\citenamefont {Klco},
  \citenamefont {Stryker},\ and\ \citenamefont {Savage}}]{Klco:2019evd}%
  \BibitemOpen
  \bibfield  {author} {\bibinfo {author} {\bibfnamefont {N.}~\bibnamefont
  {Klco}}, \bibinfo {author} {\bibfnamefont {J.~R.}\ \bibnamefont {Stryker}}, \
  and\ \bibinfo {author} {\bibfnamefont {M.~J.}\ \bibnamefont {Savage}},\
  }\href@noop {} {\  (\bibinfo {year} {2019})},\ \Eprint
  {http://arxiv.org/abs/1908.06935} {arXiv:1908.06935 [quant-ph]} \BibitemShut
  {NoStop}%
\bibitem [{\citenamefont {Davoudi}\ \emph {et~al.}(2019)\citenamefont
  {Davoudi}, \citenamefont {Hafezi}, \citenamefont {Monroe}, \citenamefont
  {Pagano}, \citenamefont {Seif},\ and\ \citenamefont
  {Shaw}}]{Davoudi:2019bhy}%
  \BibitemOpen
  \bibfield  {author} {\bibinfo {author} {\bibfnamefont {Z.}~\bibnamefont
  {Davoudi}}, \bibinfo {author} {\bibfnamefont {M.}~\bibnamefont {Hafezi}},
  \bibinfo {author} {\bibfnamefont {C.}~\bibnamefont {Monroe}}, \bibinfo
  {author} {\bibfnamefont {G.}~\bibnamefont {Pagano}}, \bibinfo {author}
  {\bibfnamefont {A.}~\bibnamefont {Seif}}, \ and\ \bibinfo {author}
  {\bibfnamefont {A.}~\bibnamefont {Shaw}},\ }\href@noop {} {\  (\bibinfo
  {year} {2019})},\ \Eprint {http://arxiv.org/abs/1908.03210} {arXiv:1908.03210
  [quant-ph]} \BibitemShut {NoStop}%
\bibitem [{\citenamefont {Jordan}\ \emph {et~al.}(2011)\citenamefont {Jordan},
  \citenamefont {Lee},\ and\ \citenamefont {Preskill}}]{Jordan:2011ci}%
  \BibitemOpen
  \bibfield  {author} {\bibinfo {author} {\bibfnamefont {S.~P.}\ \bibnamefont
  {Jordan}}, \bibinfo {author} {\bibfnamefont {K.~S.~M.}\ \bibnamefont {Lee}},
  \ and\ \bibinfo {author} {\bibfnamefont {J.}~\bibnamefont {Preskill}},\
  }\href@noop {} {\  (\bibinfo {year} {2011})},\ \bibinfo {note} {[Quant. Inf.
  Comput.14,1014(2014)]},\ \Eprint {http://arxiv.org/abs/1112.4833}
  {arXiv:1112.4833 [hep-th]} \BibitemShut {NoStop}%
\bibitem [{\citenamefont {Jordan}\ \emph {et~al.}(2014)\citenamefont {Jordan},
  \citenamefont {Lee},\ and\ \citenamefont {Preskill}}]{Jordan:2014tma}%
  \BibitemOpen
  \bibfield  {author} {\bibinfo {author} {\bibfnamefont {S.~P.}\ \bibnamefont
  {Jordan}}, \bibinfo {author} {\bibfnamefont {K.~S.~M.}\ \bibnamefont {Lee}},
  \ and\ \bibinfo {author} {\bibfnamefont {J.}~\bibnamefont {Preskill}},\
  }\href@noop {} {\  (\bibinfo {year} {2014})},\ \Eprint
  {http://arxiv.org/abs/1404.7115} {arXiv:1404.7115 [hep-th]} \BibitemShut
  {NoStop}%
\bibitem [{\citenamefont {Jordan}\ \emph {et~al.}(2017)\citenamefont {Jordan},
  \citenamefont {Krovi}, \citenamefont {Lee},\ and\ \citenamefont
  {Preskill}}]{Jordan:2017lea}%
  \BibitemOpen
  \bibfield  {author} {\bibinfo {author} {\bibfnamefont {S.~P.}\ \bibnamefont
  {Jordan}}, \bibinfo {author} {\bibfnamefont {H.}~\bibnamefont {Krovi}},
  \bibinfo {author} {\bibfnamefont {K.~S.~M.}\ \bibnamefont {Lee}}, \ and\
  \bibinfo {author} {\bibfnamefont {J.}~\bibnamefont {Preskill}},\ }\href@noop
  {} {\  (\bibinfo {year} {2017})},\ \Eprint {http://arxiv.org/abs/1703.00454}
  {arXiv:1703.00454 [quant-ph]} \BibitemShut {NoStop}%
\bibitem [{\citenamefont {Luscher}(1991)}]{Luscher:1990ux}%
  \BibitemOpen
  \bibfield  {author} {\bibinfo {author} {\bibfnamefont {M.}~\bibnamefont
  {Luscher}},\ }\href {\doibase 10.1016/0550-3213(91)90366-6} {\bibfield
  {journal} {\bibinfo  {journal} {Nucl. Phys.}\ }\textbf {\bibinfo {volume}
  {B354}},\ \bibinfo {pages} {531} (\bibinfo {year} {1991})}\BibitemShut
  {NoStop}%
\bibitem [{\citenamefont {Mueller}\ \emph {et~al.}(2019)\citenamefont
  {Mueller}, \citenamefont {Tarasov},\ and\ \citenamefont
  {Venugopalan}}]{Mueller:2019qqj}%
  \BibitemOpen
  \bibfield  {author} {\bibinfo {author} {\bibfnamefont {N.}~\bibnamefont
  {Mueller}}, \bibinfo {author} {\bibfnamefont {A.}~\bibnamefont {Tarasov}}, \
  and\ \bibinfo {author} {\bibfnamefont {R.}~\bibnamefont {Venugopalan}},\
  }\href@noop {} {\  (\bibinfo {year} {2019})},\ \Eprint
  {http://arxiv.org/abs/1908.07051} {arXiv:1908.07051 [hep-th]} \BibitemShut
  {NoStop}%
\bibitem [{\citenamefont {Collins}\ and\ \citenamefont
  {Soper}(1982)}]{Collins:1981uw}%
  \BibitemOpen
  \bibfield  {author} {\bibinfo {author} {\bibfnamefont {J.~C.}\ \bibnamefont
  {Collins}}\ and\ \bibinfo {author} {\bibfnamefont {D.~E.}\ \bibnamefont
  {Soper}},\ }\href {\doibase 10.1016/0550-3213(82)90021-9} {\bibfield
  {journal} {\bibinfo  {journal} {Nucl. Phys.}\ }\textbf {\bibinfo {volume}
  {B194}},\ \bibinfo {pages} {445} (\bibinfo {year} {1982})}\BibitemShut
  {NoStop}%
\bibitem [{\citenamefont {Jordan}\ and\ \citenamefont
  {Wigner}(1928)}]{Jordan:1928wi}%
  \BibitemOpen
  \bibfield  {author} {\bibinfo {author} {\bibfnamefont {P.}~\bibnamefont
  {Jordan}}\ and\ \bibinfo {author} {\bibfnamefont {E.~P.}\ \bibnamefont
  {Wigner}},\ }\href {\doibase 10.1007/BF01331938} {\bibfield  {journal}
  {\bibinfo  {journal} {Z. Phys.}\ }\textbf {\bibinfo {volume} {47}},\ \bibinfo
  {pages} {631} (\bibinfo {year} {1928})}\BibitemShut {NoStop}%
\bibitem [{\citenamefont {Ortiz}\ \emph {et~al.}(2001)\citenamefont {Ortiz},
  \citenamefont {Gubernatis}, \citenamefont {Knill},\ and\ \citenamefont
  {Laflamme}}]{Ortiz:2000gc}%
  \BibitemOpen
  \bibfield  {author} {\bibinfo {author} {\bibfnamefont {G.}~\bibnamefont
  {Ortiz}}, \bibinfo {author} {\bibfnamefont {J.~E.}\ \bibnamefont
  {Gubernatis}}, \bibinfo {author} {\bibfnamefont {E.}~\bibnamefont {Knill}}, \
  and\ \bibinfo {author} {\bibfnamefont {R.}~\bibnamefont {Laflamme}},\ }\href
  {\doibase 10.1103/PhysRevA.64.022319} {\bibfield  {journal} {\bibinfo
  {journal} {Phys. Rev.}\ }\textbf {\bibinfo {volume} {A64}},\ \bibinfo {pages}
  {022319} (\bibinfo {year} {2001})},\ \Eprint
  {http://arxiv.org/abs/cond-mat/0012334} {arXiv:cond-mat/0012334 [cond-mat]}
  \BibitemShut {NoStop}%
\bibitem [{\citenamefont {Engels}\ \emph {et~al.}(1982)\citenamefont {Engels},
  \citenamefont {Karsch}, \citenamefont {Satz},\ and\ \citenamefont
  {Montvay}}]{Engels:1981qx}%
  \BibitemOpen
  \bibfield  {author} {\bibinfo {author} {\bibfnamefont {J.}~\bibnamefont
  {Engels}}, \bibinfo {author} {\bibfnamefont {F.}~\bibnamefont {Karsch}},
  \bibinfo {author} {\bibfnamefont {H.}~\bibnamefont {Satz}}, \ and\ \bibinfo
  {author} {\bibfnamefont {I.}~\bibnamefont {Montvay}},\ }\href {\doibase
  10.1016/0550-3213(82)90077-3} {\bibfield  {journal} {\bibinfo  {journal}
  {Nucl. Phys.}\ }\textbf {\bibinfo {volume} {B205}},\ \bibinfo {pages} {545}
  (\bibinfo {year} {1982})}\BibitemShut {NoStop}%
\bibitem [{\citenamefont {Burgers}\ \emph {et~al.}(1988)\citenamefont
  {Burgers}, \citenamefont {Karsch}, \citenamefont {Nakamura},\ and\
  \citenamefont {Stamatescu}}]{Burgers:1987mb}%
  \BibitemOpen
  \bibfield  {author} {\bibinfo {author} {\bibfnamefont {G.}~\bibnamefont
  {Burgers}}, \bibinfo {author} {\bibfnamefont {F.}~\bibnamefont {Karsch}},
  \bibinfo {author} {\bibfnamefont {A.}~\bibnamefont {Nakamura}}, \ and\
  \bibinfo {author} {\bibfnamefont {I.~O.}\ \bibnamefont {Stamatescu}},\ }\href
  {\doibase 10.1016/0550-3213(88)90644-X} {\bibfield  {journal} {\bibinfo
  {journal} {Nucl. Phys.}\ }\textbf {\bibinfo {volume} {B304}},\ \bibinfo
  {pages} {587} (\bibinfo {year} {1988})}\BibitemShut {NoStop}%
\bibitem [{\citenamefont {Pedernales}\ \emph {et~al.}(2014)\citenamefont
  {Pedernales}, \citenamefont {{Di Candia}}, \citenamefont {Egusquiza},
  \citenamefont {Casanova},\ and\ \citenamefont
  {Solano}}]{PhysRevLett.113.020505}%
  \BibitemOpen
  \bibfield  {author} {\bibinfo {author} {\bibfnamefont {J.~S.}\ \bibnamefont
  {Pedernales}}, \bibinfo {author} {\bibfnamefont {R.}~\bibnamefont {{Di
  Candia}}}, \bibinfo {author} {\bibfnamefont {I.~L.}\ \bibnamefont
  {Egusquiza}}, \bibinfo {author} {\bibfnamefont {J.}~\bibnamefont {Casanova}},
  \ and\ \bibinfo {author} {\bibfnamefont {E.}~\bibnamefont {Solano}},\ }\href
  {\doibase 10.1103/PhysRevLett.113.020505} {\bibfield  {journal} {\bibinfo
  {journal} {Phys. Rev. Lett.}\ }\textbf {\bibinfo {volume} {113}},\ \bibinfo
  {pages} {020505} (\bibinfo {year} {2014})}\BibitemShut {NoStop}%
\bibitem [{\citenamefont {Roggero}\ and\ \citenamefont
  {Carlson}(2018)}]{Roggero:2018hrn}%
  \BibitemOpen
  \bibfield  {author} {\bibinfo {author} {\bibfnamefont {A.}~\bibnamefont
  {Roggero}}\ and\ \bibinfo {author} {\bibfnamefont {J.}~\bibnamefont
  {Carlson}},\ }\href@noop {} {\  (\bibinfo {year} {2018})},\ \Eprint
  {http://arxiv.org/abs/1804.01505} {arXiv:1804.01505 [quant-ph]} \BibitemShut
  {NoStop}%
\bibitem [{\citenamefont {Belitsky}\ \emph {et~al.}(2014)\citenamefont
  {Belitsky}, \citenamefont {M{\"u}ller},\ and\ \citenamefont
  {Ji}}]{Belitsky:2012ch}%
  \BibitemOpen
  \bibfield  {author} {\bibinfo {author} {\bibfnamefont {A.~V.}\ \bibnamefont
  {Belitsky}}, \bibinfo {author} {\bibfnamefont {D.}~\bibnamefont
  {M{\"u}ller}}, \ and\ \bibinfo {author} {\bibfnamefont {Y.}~\bibnamefont
  {Ji}},\ }\href {\doibase 10.1016/j.nuclphysb.2013.11.014} {\bibfield
  {journal} {\bibinfo  {journal} {Nucl. Phys.}\ }\textbf {\bibinfo {volume}
  {B878}},\ \bibinfo {pages} {214} (\bibinfo {year} {2014})},\ \Eprint
  {http://arxiv.org/abs/1212.6674} {arXiv:1212.6674 [hep-ph]} \BibitemShut
  {NoStop}%
\bibitem [{\citenamefont {Kaplan}\ \emph {et~al.}(2017)\citenamefont {Kaplan},
  \citenamefont {Klco},\ and\ \citenamefont {Roggero}}]{Kaplan:2017ccd}%
  \BibitemOpen
  \bibfield  {author} {\bibinfo {author} {\bibfnamefont {D.~B.}\ \bibnamefont
  {Kaplan}}, \bibinfo {author} {\bibfnamefont {N.}~\bibnamefont {Klco}}, \ and\
  \bibinfo {author} {\bibfnamefont {A.}~\bibnamefont {Roggero}},\ }\href@noop
  {} {\  (\bibinfo {year} {2017})},\ \Eprint {http://arxiv.org/abs/1709.08250}
  {arXiv:1709.08250 [quant-ph]} \BibitemShut {NoStop}%
\bibitem [{\citenamefont {Lamm}\ and\ \citenamefont
  {Lawrence}(2018)}]{Lamm:2018siq}%
  \BibitemOpen
  \bibfield  {author} {\bibinfo {author} {\bibfnamefont {H.}~\bibnamefont
  {Lamm}}\ and\ \bibinfo {author} {\bibfnamefont {S.}~\bibnamefont
  {Lawrence}},\ }\href {\doibase 10.1103/PhysRevLett.121.170501} {\bibfield
  {journal} {\bibinfo  {journal} {Phys. Rev. Lett.}\ }\textbf {\bibinfo
  {volume} {121}},\ \bibinfo {pages} {170501} (\bibinfo {year} {2018})},\
  \Eprint {http://arxiv.org/abs/1806.06649} {arXiv:1806.06649 [quant-ph]}
  \BibitemShut {NoStop}%
\bibitem [{\citenamefont {Messiah}(1969)}]{messiah1962quantum}%
  \BibitemOpen
  \bibfield  {author} {\bibinfo {author} {\bibfnamefont {A.}~\bibnamefont
  {Messiah}},\ }\href@noop {} {\emph {\bibinfo {title} {Quantum mechanics,
  volume II}}}\ (\bibinfo  {publisher} {North-Holland Publishing Company},\
  \bibinfo {year} {1969})\BibitemShut {NoStop}%
\end{thebibliography}%
\end{document}